\documentclass[a4paper,fleqn,sort&compress]{cas-sc}

\usepackage[utf8]{inputenc}
\usepackage[english]{babel}
\usepackage[square,numbers]{natbib}
\usepackage{placeins}
\usepackage{amsmath}
\usepackage{graphicx}
\usepackage{tcolorbox}
\usepackage{relsize}
\usepackage{comment}
\usepackage{tabu,multirow}
\usepackage{epsfig}
\usepackage{array}
\usepackage{amssymb}
\usepackage{latexsym}
\usepackage{xcolor}
\usepackage{subfigure}
\usepackage{soul}
\usepackage[export]{adjustbox}
\usepackage{setspace}
\usepackage{caption}
\usepackage{float}
\usepackage{tikz}
\usetikzlibrary{positioning, shapes, arrows}
\newcommand{\boldface}[1]{\boldsymbol{#1}}  % italic (slanted)

\newcommand{\bfp}{\boldface{p}}
\newcommand{\bfq}{\boldface{q}}

\newcommand{\bfx}{\boldface{x}}

\newcommand{\bfz}{\boldface{z}}

\newcommand{\bfF}{\boldface{F}}

\newcommand{\bfR}{\boldface{R}}

\newcommand{\bfnull}{\boldsymbol{0}}

\newcommand{\bfSigma}{\boldsymbol{\Sigma}}

\newcommand{\bfOmega}{\boldsymbol{\Omega}}

\newcommand{\calC}{\mathcal{C}}

\newcommand{\calH}{\mathcal{H}}

\newcommand{\calL}{\mathcal{L}}

\newcommand{\calN}{\mathcal{N}}

\newcommand{\calZ}{\mathcal{Z}}
\newcommand{\T}{^{\mathrm{T}}} % x^{T}
\newcommand{\Rset}{\mathbb{R}}

\newlength{\boxwidth}
\def\dd{\;\!\mathrm{d}}

\def\btheorem{\begin{theorem}}
\def\etheorem{\end{theorem}}
\def\blemma{\begin{lemma}}
\def\elemma{\end{lemma}}
\def\bproposition{\begin{proposition}}
\def\eproposition{\end{proposition}}
\def\bcorollary{\begin{corollary}}
\def\ecorollary{\end{corollary}}
\def\bdefinition{\begin{definition}}
\def\edefinition{\end{definition}}
\def\bexample{\begin{example}}
\def\eexample{\end{example}}
\def\bremark{\begin{remark}}
\def\eremark{\end{remark}}

\DeclareMathOperator{\tr}{tr}

\newcommand{\be}{\begin{equation*}}
\newcommand{\ee}{\end{equation*}}

\newcommand{\beq}{\begin{eqnarray*}}
\newcommand{\eeq}{\end{eqnarray*}}
\newcommand{\bem}{\begin{multline}}
\newcommand{\eem}{\end{multline}}
\newcommand{\ba}{\begin{align*}}
\newcommand{\ea}{\end{align*}}

\newcommand{\itbf}[1]{\textit{\textbf{#1}}}

\newcommand{\inner}[1]{\left\langle#1\right\rangle}

\tikzstyle{terminator} = [black, rectangle, draw, text centered, rounded corners, minimum height=2em]
\tikzstyle{process} = [black, rectangle, draw, text centered, minimum height=2em]
\tikzstyle{decision} = [black, diamond, draw, text centered, minimum height=2em]
\tikzstyle{connector} = [black, draw, -latex']
\tikzstyle{data}=[black, trapezium, draw, text centered, trapezium left angle=80, trapezium right angle=100, minimum height=2em]

\begin{document}
\let\WriteBookmarks\relax
\def\floatpagepagefraction{1}
\def\textpagefraction{.001}
\shorttitle{Phase Space Integration}
\shortauthors{Saxena et~al.}
%\begin{frontmatter}

\title [mode = title]{GNN-Assisted Phase Space Integration with Application to Atomistics}

% \author[1,3]{CV Radhakrishnan}[type=editor,
%                         auid=000,bioid=1,
%                         prefix=Sir,
%                         role=Researcher,
%                         orcid=0000-0001-7511-2910]
% \cormark[1]
% \ead{cvr_1@tug.org.in}
% \ead[url]{www.cvr.cc, cvr@sayahna.org}

% \credit{Conceptualization of this study, Methodology, Software}

% \address[1]{Elsevier B.V., Radarweg 29, 1043 NX Amsterdam, The Netherlands}

% \author[2,4]{Han Theh Thanh}[style=chinese]

% \author[2,3]{CV Rajagopal}[%
%   role=Co-ordinator,
%   suffix=Jr,
% \ead{cvr3@sayahna.org}
% \ead[URL]{www.sayahna.org}

% \credit{Data curation, Writing - Original draft preparation}

% \address[2]{Sayahna Foundation, Jagathy, Trivandrum 695014, India}

% \author%
% [1,3]
% {Rishi T.}
% \cormark[2]
% \fnmark[1,3]
% \ead{rishi@stmdocs.in}
% \ead[URL]{www.stmdocs.in}

\author[1]{Shashank Saxena}[orcid=0000-0002-5242-9103]
\author[1]{Jan-Hendrik Bastek}[orcid=0000-0003-3343-1977]
\author[1]{Miguel Spinola}[orcid=0000-0002-5180-6149]
\author[2]{Prateek Gupta}[orcid=0000-0003-3666-0257]
\ead{prgupta@am.iitd.ac.in}
\author[1]{Dennis M. Kochmann}[orcid=0000-0002-9112-6615]
\cormark[1]
\ead{dmk@ethz.ch}
\ead[URL]{mm.ethz.ch}
\address[1]{Mechanics \& Materials Laboratory, Department of Mechanical and Process Engineering, ETH Z\"urich, 8092 Z\"urich, Switzerland}
\address[2]{Department of Applied Mechanics,  Indian Institute of Technology Delhi, 110016, New Delhi, India}

\cortext[cor1]{Corresponding author}

% \fntext[fn1]{This is the first author footnote. but is common to third
% \fntext[fn2]{Another author footnote, this is a very long footnote and
%   it should be a really long footnote. But this footnote is not yet
%   sufficiently long enough to make two lines of footnote text.}

% \nonumnote{This note has no numbers. In this work we demonstrate $a_b$
%   the formation Y\_1 of a new type of polariton on the interface
%   between a cuprous oxide slab and a polystyrene micro-sphere placed

\begin{abstract}
Overcoming the time scale limitations of atomistics can be achieved by switching from the state-space representation of Molecular Dynamics (MD) to a statistical-mechanics-based representation in phase space, where approximations such as maximum-entropy or Gaussian phase packets (GPP) evolve the atomistic ensemble in a time-coarsened fashion. In practice, this requires the computation of expensive high-dimensional integrals over all of phase space of an atomistic ensemble. This, in turn, is commonly accomplished efficiently by low-order numerical quadrature.
We show that numerical quadrature in this context, unfortunately, comes with a set of inherent problems, which corrupt the accuracy of simulations---especially when dealing with crystal lattices with imperfections. As a remedy, we demonstrate that Graph Neural Networks, trained on Monte-Carlo data, can serve as a replacement for commonly used numerical quadrature rules, overcoming their deficiencies and significantly improving the accuracy. This is showcased by three benchmarks: the thermal expansion of copper, the martensitic phase transition of iron, and the energy of grain boundaries. We illustrate the benefits of the proposed technique over classically used third- and fifth-order Gaussian quadrature, we highlight the impact on time-coarsened atomistic predictions, and we discuss the computational efficiency. The latter is of general importance when performing frequent evaluation of phase space or other high-dimensional integrals, which is why the proposed framework promises applications beyond the scope of atomistics.
\end{abstract}


\maketitle

\section{Introduction} \label{Intro}
Evaluating integrals in high-dimensional spaces is required in many engineering and scientific applications. High dimensionality, sometimes involving millions or billions of dimensions, is usually an outcome of mathematical modeling of complex systems. For example, high-dimensional spaces in quantum physics result from solving the multi-particle Schr\"odinger equation \cite{beylkin2005algorithms,lanzara2019fast}, where every particle adds to the dimension of the problem. Analogously in the study of dynamical systems\cite{j2007statistical,zubarev1989nonequilibrium,frenkel2001understanding}, the evolution of an ensemble of degrees of freedom is studied in the system's phase space. 
% For instance, in classical statistical physics, ensemble of atoms/molecules/particles is studied in the phase space defined by positions and momenta of the constituent particles. 
Considering, e.g., an ensemble of atoms, molecules, or particles in three-dimensional (3D) physical space, each particle---characterized by its position and momentum vectors---adds six dimensions to the problem. Identifying reaction pathways in chemistry and molecular biology \cite{lotstedt2006dimensional, lecca2019theoretical} is another example, where the chemical master equation (CME) is used to model the probabilities of discrete molecular species in the system  \cite{sjoberg2005numerical}, each of which adds one dimension to the problem. Since there are thousands of proteins in a biological system, the number of dimensions of the system can be intractably large. Similar high-dimensional Fokker-Plank equations are used to study the dynamics of polymeric fluids \cite{venkiteswaran2005qmc,venkiteswaran2005quasi}, where each molecule in a polymer chain adds a dimension to the system. Besides basic science and engineering, high-dimensional integrals are abundant in financial mathematics. The price of financial derivatives is evaluated as an expectation value over a multi-dimensional space of hundreds of random variables, which are the sources of uncertainty captured by a financial model \cite{holtz2010sparse, sullivan2015introduction, paskov1996faster}.
% Finding the payoff of so-called Asian options \cite{lapeyre2001competitive}, whose price depends on the temporal evolution of a stochastic process over a given period of time, also involves high-dimensional integrals. 
Other relevant areas include dynamic reinforcement learning \cite{kim2020hamilton,engel2005reinforcement} and data mining techniques \cite{garcke2010data}. 

A key challenge in mechanics that fits into this context of high-dimensional spaces is the computational modeling of atomic ensembles at finite temperature.
In its nascent stages, Molecular Dynamics (MD) was used to simulate an ensemble of hard spheres interacting via an interatomic potential~\cite{alder1959studies}, shortly followed by more complex potentials for studying radiation damage \cite{gibson1960dynamics} and dilute gases \cite{rahman1964correlations}. Coupled with thermostats \cite{nose1984unified} and barostats \cite{parrinello1981polymorphic} to capture ergodic thermal equilibrium distributions \cite{tadmor2011modeling}, MD has become a state-of-the-art technique for studying the mechanics and thermodynamics, physics and chemistry of materials at nanometer length scales. However, the maximum time scales accessible by MD are limited to microseconds on present-day computers, which prevents MD from simulating, e.g., many slow-scale transport processes and associated physical and chemical properties \cite{ziman2001electrons}. Therefore, many attempts have been made to accelerate MD simulations \cite{frenkel2001understanding,voter1997method, voter1998parallel, tadmor2013finite, dongare2014quasi, Gromacs}. In the following, we focus on statistical-mechanics-based techniques, which aim to approximate the probability distribution of an atomic ensemble at finite temperature. This class of methods includes \textit{diffusive molecular dynamics} (DMD) \cite{li2011diffusive}, \textit{maximum-enttropy}  (\textit{max-ent}) \cite{kulkarni2008variational}, and \textit{Gaussian phase packets} (GPP) \cite{gupta2021nonequilibrium} formulations. In the quasistatic limit, all those formulations converge to the same problem of iteratively solving a set of coupled nonlinear equations to satisfy vanishing phase-averaged physical and thermal forces at each atomic site. The phase averaging of forces and potential energy is performed by integrating over the entire phase space.

All of the aforementioned applications involve the numerical computation of high-dimensional integrals and suffer from the \textit{curse of dimensionality} \cite{COD,taylor2019applications}, which states that the number of samples needed to approximate integrals with a comparable accuracy increases exponentially with the dimension of the problem.  Monte-Carlo methods have been a natural choice for such problems because of their dimension-independent convergence of order $O(n^{-1/2})$, where $n$ equals the number of sampling points. However, it does not provide an accurate error estimate, and studies \cite{sloan2004does} suggest that the coefficient multiplying the error bound might depend on the problem's dimensionality. More efficient schemes with better error estimates and convergence rates, such as Quasi Monte-Carlo (QMC) \cite{morokoff1995quasi} and Randomized Quasi Monte-Carlo (RQMC) \cite{tuffin2004randomization}, involve sampling over deterministic and randomly shifted points, respectively, from low-discrepancy (e.g., Sobol) sequences \cite{niederreiter1992random}, and also have a dimension-independent convergence order. The \textit{curse of dimensionality} can be avoided if the number of sampling points required to approximate an integral does not increase exponentially with the dimension. This phenomenon is known as \textit{tractability} in complexity theory. Many studies have therefore focused on the tractability of numerical integration schemes \cite{sloan2004finite,mhaskar2004tractability,wang2003strong}. Brownian bridge construction \cite{caflisch1997valuation} is a common technique in financial mathematics to reduce the effective dimension \cite{wang2005high} of a problem by expressing a function of multiple variables as a sum of multiple functions depending on chosen sets of those variables. This is similar to the technique used in atomic ensembles with a short-range interaction, where the total energy of a system can be written as a sum of the energies of individual atoms, which depend on the neighbors of the atoms \cite{tembhekar2018fully}. Other attempts to efficiently compute high-dimensional integrals involve the use of sparse grids \cite{holtz2010sparse,griebel2005sparse} in the phase space and moment approaches \cite{bertsimas2006multivariate}, which show the asymptotic convergence of the upper and lower bounds of the integral with increasing order of moments. Finally, works in the late 1990's \cite{kainen1997utilizing,donoho2000high} started suggesting that probability distributions in high dimensions have random points clustered around a lower-dimensional geometric shape, and hence constructing such data points for machine learning is a simple alternative \cite{gorban2018blessing}. This property was coined as \textit{blessing of dimensionality}. However, finding such clusters in high dimensions is difficult \citet{pestov2013k}, and hence the curse and blessing of dimensionality are two sides of the same coin. 

For efficient calculations of phase space integrals, current techniques in \textit{time-coarsened atomic simulations} involve the use of multivariable Gaussian quadrature rules \cite{stroud1971approximate}. The sampling points for such quadrature rules require sequential shifting of the atoms of a cluster along the Cartesian axes and averaging over the energies of the individual configurations with appropriate weights. The third-order quadrature rule is an efficient choice, leading to as much as a ten-fold reduction in computation times compared to MD, without any practical loss of accuracy, in the example of calculating surface properties of metals~\cite{saxena2022fast}. Higher-order quadrature rules are more accurate but are computationally costly and hence reduce the overall computational efficiency of simulations. We later show that using a fifth-order quadrature rule already gives computational times comparable to that of MD. We further show that using such Gaussian quadrature rules makes the energy computation inobjective with respect to proper rotations in the $SO(3)$ orthogonal group and leads to incorrect jumps in the energy profile along a deformation path when dealing with large deformations. Such a non-physical behaviour limits the use of these time-coarsened atomistic techniques to simplistic scenarios. As a remedy, we here present the use of an $E(3)$ equivariant graph neural network (GNN) \cite{batzner20223} to learn the phase averages of the energy of an atomistic ensemble and its derivatives, using highly accurate Monte-Carlo data for training. The GNN model, once trained offline, can be leveraged to simulate atomistic ensembles at finite temperature at a significantly improved accuracy-to-cost ratio for diverse applications including amorphous microstructures, phase transitions, and fracture. 

The remainder of this contribution is structured as follows. Section \ref{review of quadrature rules} gives a brief overview of phase space averaging using Gaussian quadrature rules in the context of the GPP formalism for time-coarsened atomistics. Section \ref{List of problems} discusses the three prime issues when using quadrature rules: the frame dependence of the integrated energy, discontinuities in the energy during deformation, and the loss of accuracy. In Section \ref{GNN architecture} we describe the neural network architecture used to address the aforementioned challenges, adapting the approach in \citet{batzner20223} for our purposes. There, we also provide details of the Monte-Carlo integration scheme used to generate the training data. Section \ref{results} presents results from the application of the trained GNN models to selected benchmark applications comparison with MD results obtained using the Large-scale Atomic/Molecular Massively Parallel Simulator (LAMMPS) \cite{LAMMPS}. Finally, we conclude this study in Section \ref{conclusions} along with a brief discussion of open challenges and possible extensions.

\section{Approximating Phase Space Averages in Time-Coarsened Atomistics} 
\label{review of quadrature rules}

We begin by briefly reviewing the framework of time-coarsened atomistics at finite temperature, from which emerges the need to evaluate high-dimensional phase space integrals of the potential energy and of forces in the atomic ensembles. More specifically, we consider the statistical-mechanics-based DMD \cite{li2011diffusive}, max-ent \cite{kulkarni2008variational}, and GPP \cite{gupta2021nonequilibrium} frameworks, which aim at computing the statistically averaged mean motion of an ensemble of atoms over considerably longer time scales than those accessible by MD. To this end, those techniques aim to replace the positions $\bfq=\{\bfq_i(t) : i=1,\ldots,N\}$ and momenta  $\bfp=\{\bfp_i(t) : i=1,\ldots,N\}$ of all $N$ atoms in the ensemble by the time-averaged mean positions $\bar\bfq=\{\bar\bfq_i : i=1,\ldots,N\}$, mean momenta  $\bar\bfp=\{\bar\bfp_i : i=1,\ldots,N\}$, and the corresponding statistical variances in positions and momenta, all of which evolve at significantly slower rates than $\bfq$ and $\bfp$. For an ergodic system in thermodynamic equilibrium, the link between the two representations is made through a probability distribution function $f (\bfz ,t)$, parameterized by the positions and momenta, which for simplicity we combine into the phase space coordinate $\bfz=\left(\bfp(t),\bfq(t)\right)  \in \mathbb{R}^{6N}$. Given that the ergodic assumption holds, time averages can be replaced by phase space averages, so that we may write $\bar q=\inner q$ and $\bar p=\inner p$, where the phase average is defined as the phase space integral
\begin{equation}
    \left\langle{\cdot}\right\rangle = \frac{1}{\mathcal{Z}(t)}\int\displaylimits_{\mathbb{R}^{6N}}\left(\cdot\right) f(\bfz, t) \dd\bfz
\end{equation}
with a partition function $\calZ(t)$, ensuring that $\inner{1}=1$. Based positions and momenta, other relevant thermodynamic quantities of interest at time $t$ (such as atomic energies) can be obtained by the above phase averaging with respect to the probability distribution function.

Since the three above frameworks (DMD, max-ent, and GPP) all converge to the same equations of motion in the quasistatic limit, we here give a brief summary only of the (GPP) formulation of \citet{gupta2021nonequilibrium}. Following \citet{ma1993approximate}, the GPP formulation posits a multivariate Gaussian form of the probability distribution function for the phase space coordinate $\bfz=\left(\bfp,\bfq\right)$. The covariance between positions and momenta of atoms are contained in the covariance matrix
\begin{equation}
\boldsymbol{\Sigma} = \langle (\bfz - \bar{\bfz} )\otimes (\bfz - \bar{\bfz} ) \rangle \in \mathbb{R}^{6N\times 6N}.
\end{equation}
%  The overbar $(\bar{\cdot})$ in the above equation denotes the phase average defined as the phase space integral with respect to $f(\bfz, t)$
% \begin{equation}
%     \bar{\cdot} = \left\langle{\cdot}\right\rangle = \frac{1}{\mathcal{Z}(t)}\int\displaylimits_{\mathbb{R}^{6N}}\left(\cdot\right) f(\bfz, t) d\bfz.
% \end{equation}
In the most general case, the covariances between positions and momenta of different atoms
are non-zero. To render the problem computationally tractable, an interatomic independent assumption (i.e., $\bfSigma_{ij}=0$ for $i\neq j$) is employed to solve for the equilibrium configuration of a system approximated as an Einstein solid, while the interatomic correlations responsible for nonequilibrium irreversible thermal transport must be modeled separately. Consequently, we limit ourselves to independent Gaussian phase packets, which implies
\begin{equation}
    f(\bfz ,t) = \prod_{i=1}^N f_i(\bfz _i,t)
    \qquad\text{with}\qquad
    f_i(\bfz _i,t) = \frac{1}{\calZ _i(t)} \exp\left[ -\frac{1}{2}\left(\bfz _i - \bar{\bfz}_i(t)\right)\T \boldsymbol{\Sigma}_i^{-1}(t) \left(\bfz _i - \bar{\bfz}_i(t)\right)\right],
    \label{independent GPP}
\end{equation}
where $\bfz_i=\left(\bfp_i(t),\bfq_i(t)\right)  \in \mathbb{R}^{6}$ and $\calZ _i(t)$ are, respectively, the phase space coordinate and partition function of the $i^{th}$ atom at time $t$.

Further assuming a hyperspherical shape of the atomic distribution function $f _i$ in six dimensions leads to vanishing correlations of positions and momenta in different directions. The only non-zero terms thus remaining in the covariance matrix are (with $\tr(\cdot)$ denoting the trace of a matrix)
\begin{equation}
    \Omega_i = \frac{1}{3}\text{tr}\left( \boldsymbol{\Sigma}_i^{(\bfp ,\,\bfp )}\right), \qquad \Sigma_i = \frac{1}{3}\text{tr}\left( \boldsymbol{\Sigma}_i^{(\bfq ,\,\bfq )}\right), \qquad\mathrm{and}\qquad \beta_i = \frac{1}{3}\text{tr}\left( \boldsymbol{\Sigma}_i^{(\bfp ,\,\bfq )}\right),
\end{equation}
where we defined the covariance matrix, consisting of diagonal block matrices, as
\begin{equation}
 \bfSigma_{i} = \left(\begin{matrix}
                    \bfSigma^{(\bfp,\bfp)}_{i} &  \bfSigma^{(\bfp,\bfq)}_{i} \\
                     \bfSigma^{(\bfq,\bfp)}_{i} &  \bfSigma^{(\bfq,\bfq)}_{i}
                       \end{matrix}
\right).
\label{eq: lumped_covariance}
\end{equation}

The set of parameters to be solved for each atom $i=1,\ldots,N$ has become $\left( \bar\bfp _i,\bar\bfq_i,\Omega_i,\Sigma_i,\beta_i\right) $. In the quasistatic limit, mean momenta $\bar{\bfp }_i$ and mean thermal momenta $\beta_i$ vanish for every atom \cite{gupta2021nonequilibrium}, while information about the momentum variances $\Omega=\{\Omega_i:i=1,\ldots,N\}$ is obtained from the type of thermodynamic process assumed to bring the system to equilibrium (for an isothermal process, $\Omega_i = m k_B T$ holds for every atom $i$ in a system at a constant temperature~$T$; for an isentropic process, $\Omega_i \Sigma_i = \text{const.}$ for each atom $i$). Consequently, we are left with four scalar equations in the quasistatic limit:
% Their evolution over time is obtained from the phase-averaged equations of motion, which result from inserting the above into Liouville's equation \cite{gupta2021nonequilibrium}:
% \be
% \begin{split}
% &\frac{\text{d} \bar{\bfq }_i}{\text{d}t} = \frac{\langle \bfp _i \rangle }{m_i},\qquad  \frac{\text{d} \bar{\bfp }_i}{\text{d}t} = \langle \itbf{F}_i  \rangle, \\
%   & \frac{\text{d} \Omega_i}{\text{d}t} = \frac{ \langle \itbf{F}_i(\bfq ) \cdot (\bfp  - \bar{\bfp })  \rangle }{3}, \\
%   &\frac{\text{d} \Sigma_i}{\text{d}t} = \frac{2\beta_i}{m_i},\\
%   &\frac{\text{d} \beta_i}{\text{d}t} = \frac{\Omega_i}{m_i} + \frac{ \langle \itbf{F}_i(\bfq ) \cdot (\bfq  - \bar{\bfq })  \rangle }{3},
% \end{split}
% \ee
% where $\bfF_i$ denotes the net force acting on atom $i$ having mass $m_i$.
\begin{subequations}
\label{EOM quasistatic}
\begin{align}
\langle \itbf{F}_i \rangle & = \bfnull\qquad \mathrm{and}\\
\frac{\Omega_i}{m_i} + \frac{ \langle \itbf{F}_i(\bfq ) \cdot (\bfq  - \bar{\bfq })  \rangle }{3} & = 0,\label{EOM2}
\end{align}
\end{subequations}
whose solution is the set of average positions $\bar\bfq=\{ \bar\bfq _i: i=1,\ldots,N \}$ and position variances $\Sigma=\{\Sigma_i : i=1,\ldots, N\}$ for all atoms in equilibrium. Importantly, this admits decoupling the phase dynamics of thermal vibrations from the slow mean motion of atoms, which is essential towards our objective of studying equilibrium properties at finite temperature. Rather than resolving atomic motion at the femtosecond level, this approach tracks the effective atomic parameters $\left( \bar\bfp _i,\bar\bfq_i,\Omega_i,\Sigma_i,\beta_i\right) $ over time (at significantly larger time scales than required for $\left(\bfp _i,\bfq_i\right) $ in MD) and, in the quasistatic limit, reduces to a set of equilibrium equations to be solved for the aforementioned effective parameters.

% The entropy of the system is obtained using Boltzmann's expression,
% \be
%     S = -k_B \langle \ln f \rangle = \sum_{i=1}^N \Bigg[S_{0_i} - 3k_B \ln h + \frac{3k_B}{2} \ln(\Omega_i \Sigma_i) \Bigg],
% \ee
% where $S_{0_i} = 3 k_B [1+\text{ln}(2\pi)] - \text{ln}(N!)/N$, $h$ is Planck's constant, and $k_B$ Boltzmann's constant. Therefore, to simulate, e.g., an isentropic process, condition $\Omega_i \Sigma_i = \text{const.}$ for each atom $i$ complements Eqs.~\eqref{EOM quasistatic}. For an isothermal process, $\Omega_i = m k_B T$ holds for every atom $i$ in a system at a constant temperature~$T$.

% In this work, we will focus on isothermal simulations, for which the condition $\Omega_i = m k_B T$ holds for every atom $i$ in a system at a constant temperature~$T$.

Substituting $\Sigma_i = k_B T / m \omega_i^2 $ recovers the DMD \cite{li2011diffusive} and \textit{max-ent}\cite{kulkarni2008variational} frameworks, where $\omega_i$ is the \textit{Einstein} frequency for the vibration of atom $i$ about its mean position. The system of equations in \eqref{EOM quasistatic} can also be interpreted as stationarity conditions, aiming to find the minimizer of the Helmholtz free energy of the system, defined by 
\begin{equation}
    \mathcal{F}(\bar{\bfq },\Omega,\Sigma) = E (\bar{\bfq },\Omega,\Sigma) - \sum_{i=1}^N \frac{\Omega_i S_i}{k_B m_i},
    \label{Helmholtz free energy}
\end{equation}
where $S_i = -k_B \langle \ln f_i \rangle $ is the Boltzmann entropy of the $i^{th}$ atom, and we introduced the internal energy of the system as the average total Hamiltonian:
\begin{equation}\label{eq:E}
    E (\bar{\bfq },\bfOmega,\Sigma) =  
    % \langle \mathcal{H} \rangle = 
    \sum_{i=1}^N \frac{3\Omega_i}{2 m_i} + \langle V_i(\bfq ) \rangle.
\end{equation}
Here, $V(\bfq )$ is the total potential energy of the system, typically defined via interatomic potentials (as in this study). 
As the interatomic potentials involve, in general, intricate multi-body functions of atomic positions, evaluating the integrals required for the phase average of $V(\bfq )$ and its derivatives is not possible analytically. Instead, the state of the art is to use numerical Gaussian quadrature, introducing approximations of the type
\begin{equation}
 \langle V_i(\bfq) \rangle = \int_{\Gamma} V_i(\bfq) \prod^{n_N}_{j=1} \text{exp}\left[ -\frac{ \vert \bfq_{j} - \bar{\bfq}_{j} \vert^2}{2 \Sigma_{j} }  \right] \text{d}\bfq_{j}  \ \approx \ \left( \frac{1}{ \sqrt{  \pi}} \right)^{3{n_N}} \sum_{p=1}^{N_{QP}} W_P V_i(\bfq_P), \label{Quadrature}
\end{equation}
where ${n_N}$ is the number of atoms in the $i^{th}$ cluster (i.e., atom $i$ and its interacting neighbors based on the potential cut-off), $N_{QP}$ is the number of quadrature points, and $W_P$ is the weight of the $P^{\mathrm{th}}$ quadrature point at $\bfq_P=\bar\bfq+ \sqrt{2 \Sigma} \bfx_P$ with $\bfx_P\in\Rset^{3n_N}$ in 3D. A list of sampling weights and quadrature point locations for different orders was provided by \citet{stroud1971approximate}. For the reader's convenience, we reproduce in Tab.~\ref{Q3 table} a list of quadrature points $\bfx_P$ and weights $W_P$ for the most commonly used third-order Gaussian quadrature rule~\cite{kulkarni2007coarse, ariza2012hotqc, mendez2018diffusive}. For integrating over $n$ scalar variables, this quadrature requires sampling over $N_{QP}=2n$ points, for which each atom of the cluster is perturbed to the left and right of its mean position along the Cartesian axes by $\pm\sqrt{r}$. The more accurate fifth-order quadrature rule requires sampling over $N_{QP}=2n^2 + 1$ points. Therefore, for an interatomic potential that captures many-body interactions of a large number of atomic neighbors, the fifth-order quadrature becomes slightly more accurate but comes with a significant computational overhead. 

\begin{table}[h!]
\centering
\begin{tabular}{| m{1.0cm} | m{1.0cm} |  m{0.8cm} m{0.8cm} m{0.8cm} m{0.8cm}   | m{1.0cm} | m{1.0cm} |  m{0.8cm} m{0.8cm} m{0.8cm} m{0.8cm}  |  }  \hline
 $P$ & $W_P$ & $x_1$ & $x_2$ & $\cdots$ & $x_n$ & $P$ & $W_P$ & $x_1$ & $x_2$ & $\cdots$ & $x_n$ \\ \hline
 1 & $\pi^{r}/2n$ & $\sqrt{r}$ & 0 & 0 & 0 & $n+1$ & $\pi^{r}/2n$ & $-\sqrt{r}$ & 0 & 0 & 0 \\ \hline
 2 & $\pi^{r}/2n$ & 0 & $\sqrt{r}$ & 0 & 0 & $n+2$ & $\pi^{r}/2n$ & 0 & $-\sqrt{r}$ & 0 & 0 \\ \hline
 $\cdots$ & $\cdots$ & $\cdots$ &  &  &  & $\cdots$ & $\cdots$ & $\cdots$ &  &  &  \\ \hline
 $n$ & $\pi^{r}/2n$ & 0 & 0 & 0 & $\sqrt{r}$ & $2n$ & $\pi^{r}/2n$ & 0 & 0 & 0 & $-\sqrt{r}$ \\ \hline
\end{tabular}
\captionof{table}{Quadrature weights and sampling points for the third-order Gaussian quadrature for $n$ scalar variables \cite{stroud1971approximate} ($r=\frac{n}{2}$). The scalar perturbations $\{x_1,\ldots,x_n\}$ must be applied independently in each of the three Cartesian directions.}
\label{Q3 table}
\end{table}

% We use the third-order Gaussian quadrature for all presented results. If required, the accuracy may be increased by using higher-order (e.g., the fifth-order) quadrature rules. 
 
% Eq.~\eqref{Helmholtz free energy} represents a convenient way to calculate the Helmholtz free energy $\mathcal{F}$ from our GPP-based simulations, which is beneficial since $\mathcal{F}$ is the relevant thermodynamic potential of interest for finite-temperature calculations with imposed strains (and considerably harder to extract from MD, as explained in Section~\ref{methodology}). For a detailed description of the relevant Legendre transforms of the internal energy corresponding to the natural control variables for the thermodynamic process, see, e.g., \citep{alberty2001use}. In the following, we refer to the Helmholtz free energy as `\textit{free energy}' for conciseness.

\section{Problems with Gaussian Quadrature Rules} 
\label{List of problems}

The Gaussian quadrature rules discussed above come with detrimental shortcomings, which limits the accuracy of statistical-mechanics-based atomistic simulations, especially those involving complex crystal lattices or those with defects. We here discuss the mathematical origin of those problems and also present simple examples, in which they can be easily identified. Each of the subsequent subsections is dedicated to a separate problem. We begin in Section \ref{Objectivity issue} by discussing the frame dependence of phase-averaged energy (i.e., the inobjectivity of the phase space average) with respect to rotations, followed by energy discontinuities due to neighborhood changes under large deformations in Section \ref{Discontinuity issue}. Finally Section \ref{Accuracy issue} presents an example of an interatomic potential, where third-order quadrature is simply insufficient to capture atomic interactions. In each of the examples, we compare results obtained using the third- and fifth-order quadrature rules to those obtained from MD simulations.

\subsection{Inobjectivity with respect to proper rotations} 
\label{Objectivity issue}

The principle of \textit{material frame indifference} (or \textit{objectivity)}, which is well-known in continuum mechanics~\cite{tadmor2011modeling}, states that the internal energy density of a continuum material should be invariant to any rotation $\boldsymbol{R}\in SO(3)$ of the material. Although relevant even for atomistic scenarios, this principle is not commonly a concern in MD-based studies, because interatomic potentials (based on relative atomic positions) are usually by definition frame-indifferent. Moreover, MD simulations involve initializing atomic velocities along random directions, evolving them and the atomic positions according to Hamiltonian-based interatomic interactions and time averaging over $O(10^4)$ or even more time steps to record average thermodynamic quantities. Over such a large number of timesteps, a large number of instantaneous snapshots of the ensemble is sampled, so that initializing ensembles with different initial orientations has only a negligible effect. Therefore, time-averaged MD quantities (such as the internal energy, which is the phase-averaged Hamiltonian) are typically frame-indifferent. However, we show here that this is not the case for phase space-averaged quantities when using Gaussian quadrature rules. For simplicity, we restrict our example to 2D rotations, though the analogous applies in 3D.

\begin{figure}[h!]
\centering
\includegraphics[width = \textwidth]{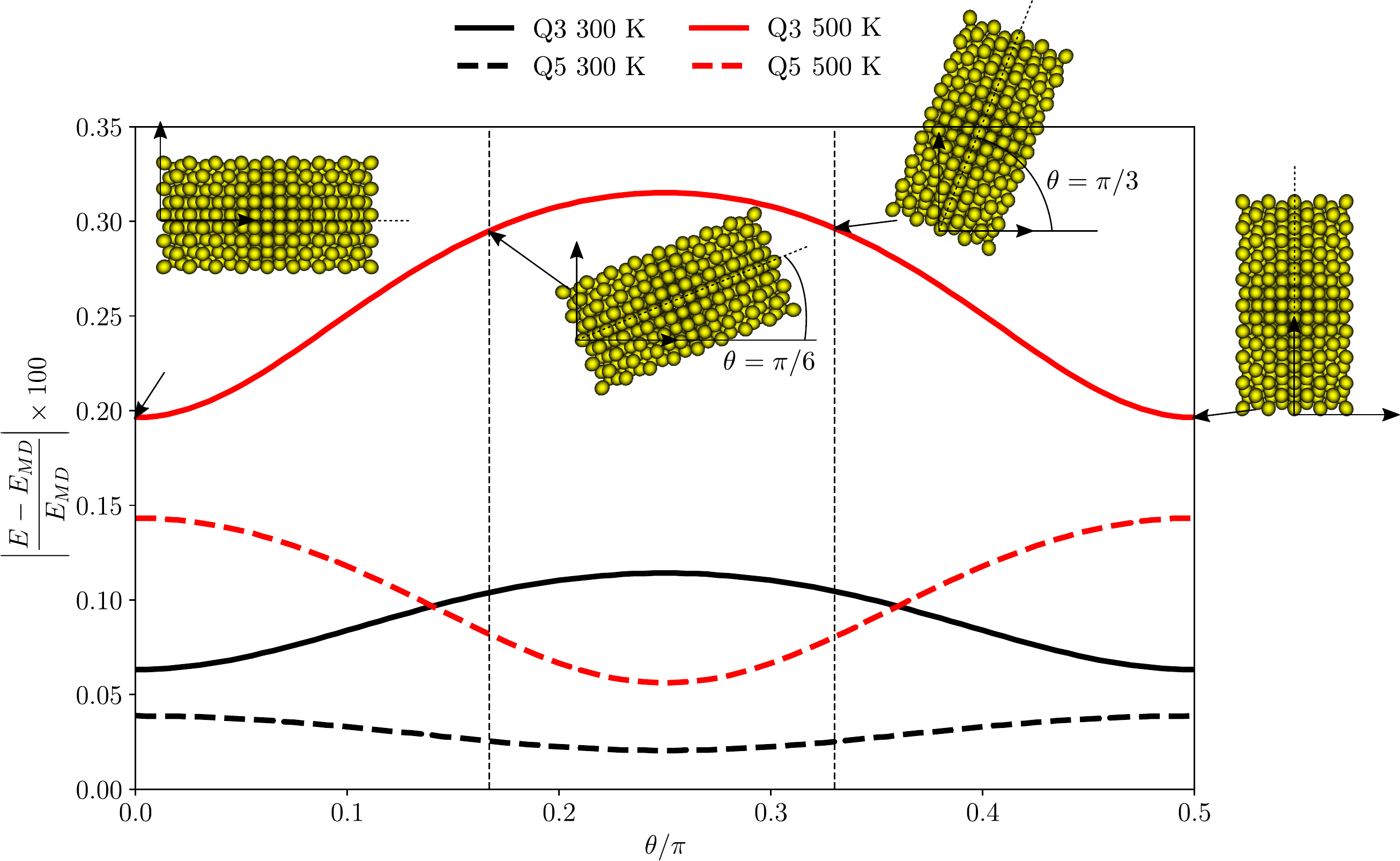}
\caption{Relative error of the relaxed internal energy for bulk copper, compared against the corresponding values from MD, as a function of the lattice rotation angle $\theta$ for third- and fifth-order quadrature rules in the GPP phase space averages. Black and red colors correspond to 300~K and 500~K, respectively. Insets show schematics of a pure copper single-crystal slab rotated by $\theta = 0,~\pi/6,~\pi/3,$~and $\pi/2$ with respect to the horizontal $x$-axis.
% \PG{Consider annotating within the plot for clarity} \DK{The sketch on the right looks a bit like my sons drew it. ;-) I would make the atoms smaller and draw some more atoms (it's a large lattice) -- instead of making them pink and the central one purple, I would use solid color for the original and open circles in a different color for the rotated ones, and then the central atom gets the solid color plus the circle around it for clarity. }  
}\label{fig:Cu_EFS_rotation}
\end{figure}

We consider a copper single-crystal, whose $\{110\}$ plane coincides with the $x$-$y$-plane, resulting in an infinite periodic lattice of $\calC_4$-symmetry in the bulk. If this crystal lattice is rotated by an arbitrary angle, its mean potential energy $\langle V\rangle$ and the internal energy $E=\inner{\calH}$ at an arbitrary temperature is unaffected by the rotation, hence confirming frame indifference. The values of $\langle V\rangle$ and $E$ at temperatures of 300~K and 500~K are obtained from MD, using a simulation cell of $10\times10\times10$ unit cells, consisting of 4000 atoms and relaxing the average atomic energies with the isentropic (\textit{NPH}) ensemble and a Langevin thermostat. Alternatively, we compute $E$ at 300~K and 500~K from \eqref{eq:E}, using the GPP framework introduced in Section~\ref{review of quadrature rules}. To check for frame indifference, we rotate the lattice in the $x$-$y$-plane in 100 steps from $\theta=0$ to $\pi/2$ and relax each crystal at fixed temperature, using the third- and fifth-order quadrature rules for the phase space integrals.
% For simplicity, we illustrate the effect of 2D rotations in the $x-y$ plane on the internal energy $(E=\langle \calH \rangle )$ per-atom for a bulk FCC pure copper crystal at 300~K and 500~K. Because the $x-y$ projection of an FCC lattice is $\calC_4$ symmetric, different atomic clusters are generated by rotating a given FCC lattice in 100 steps from $\theta=0$ to $\pi/2$ and then relaxing each cluster individually at the two temperatures using the third and fifth order quadrature rules for each. To compare the results with MD, we initialize a simulation cell of $10\times10\times10$ unit cells consisting of 4000 atoms and obtain the relaxed average energies using the isoenthalpic-isobaric ($NPH$) ensemble and a Langevin thermostat. 
Fig.~\ref{fig:Cu_EFS_rotation}(a) shows the relative error in the computed internal energy $E$ (comparing the GPP result of the rotated crystal to the MD reference value) as a function of the rotation angle~$\theta$. Results indicate significant errors in the GPP-computed values and, more importantly, strong variations of the energy with the crystal orientation---hence indicating a lack of material frame indifference.
% The relative error in energy is expressed as the percentage difference from the MD value at the corresponding temperature (which is independent of the rotation angle).

The cause of the observed variations lies in the quadrature rules, whose perturbations $\bfx_P$ (tabulated in Tab.~\ref{Q3 table}) are typically applied along the same directions defined by the Cartesian axes, irrespective of the orientation of the crystal lattice. This leads to different average internal energy values for differently rotated samples. Assume that a rotation matrix $\boldsymbol{R}\in SO(3)$ maps the set of mean atomic positions $\{ \bar{\bfq}_1,\ldots,\bar \bfq_N \}$ to $ \{ \bar{\bfq}'_1,\ldots,\bar\bfq'_N  \}=\{ \bfR\bar{\bfq}_1,\ldots,\bfR\bar \bfq_N \}$. Using Eq.~\eqref{Quadrature} without the quadrature approximation (and denoting by $\bfR\bar{\bfq}$ the vector of all rotated positions, in an abuse of notation), the average potential energy of the $i^{th}$ atom in the rotated crystal follows as 
\begin{equation}
 \langle V_i \rangle( \boldsymbol{R} \bar{\bfq},\Sigma) = \int_{\Gamma} V_i( 
\bfR\bfq) \prod^{n_N}_{j=1} \text{exp}\left[ -\frac{ \vert \bfR\bfq_{j} - \boldsymbol{R}\bar{\bfq}_{j} \vert^2}{2 \Sigma_{j} }  \right]  \text{d}\bfR\bfq_{j} \approx \ \left( \frac{1}{ \sqrt{  \pi}} \right)^{3{n_N}} \sum_{p=1}^{N_{QP}} W_P V_i(\bfR\bfq_P)    \label{true integral objectivity}
\end{equation}
% where $\bfq_j^{'} = \boldsymbol{R}^T \bfq_j $. Given that 
since the potential energy depends only on the interatomic distances $(V_i(\bfR\bfq)=V_i(\bfq))$ and $\boldsymbol{R}$ is a pure rotation, so 
\begin{equation}
\langle V_i \rangle(\boldsymbol{R}\bar{\bfq},\Sigma) =  \langle V_i \rangle(\bar{\bfq},\Sigma).
\end{equation}
However, when applying the quadrature rule in \eqref{Quadrature} for fixed quadrature perturbations $\bfx_P$, the approximated average energy (denoted by $\langle V_i \rangle^h$) is evaluated as
\begin{equation}
   \langle V_i \rangle^h(\boldsymbol{R}\bar{\bfq},\Sigma) = \left( \frac{1}{ \sqrt{  \pi}} \right)^{3{n_N}} \sum_{p=1}^{N_{QP}} W_P V( \boldsymbol{R} \bar{\bfq} + \sqrt{2\Sigma} \bfx_P ) \neq \left( \frac{1}{ \sqrt{  \pi}} \right)^{3{n_N}} \sum_{p=1}^{N_{QP}} W_P V(  \bar{\bfq} + \sqrt{2\Sigma} \bfx_P ).
\end{equation}
% The above inequality shows that rotating an isolated ensemble of atoms in space changes the internal energy of the ensemble, highlighting the non-objective nature of phase space integrals approximated by a numerical quadrature.
One may argue that rotating the quadrature perturbations $\bfx_P$ with the same rotation matrix such that $\bfx_P^{'} = \boldsymbol{R} \bfx_P$ can be a simple solution to make the phase-averaged energies invariant to rotations. However, this solution is physically infeasible for general cases. Information of crystal rotations during large deformations is physically meaningful only for perfect atomic neighborhoods. In the vicinity of lattice defects (such as grain boundaries and dislocations) there is no unique optimal choice for the set of Cartesian axes defining the quadrature perturbations. Moreover, the crystal symmetry directions for bulk atomic neighborhoods also change if the crystal undergoes phase transitions, as we show in Section~\ref{results 2}. An extreme case of the aforementioned problem are amorphous solids, where the task of identifying specific perturbation directions is impossible. This, in summary, shows that conventional Gauss quadrature rules applied to phase space averages in \eqref{Quadrature} render the computed energetic quantities (and all derived quantities such as atomic forces) inobjective or frame-dependent.

\subsection{Discontinuity at the cutoff radius} 
\label{Discontinuity issue}

\begin{figure}[!tb]
\centering
\includegraphics[width = 0.85\textwidth]{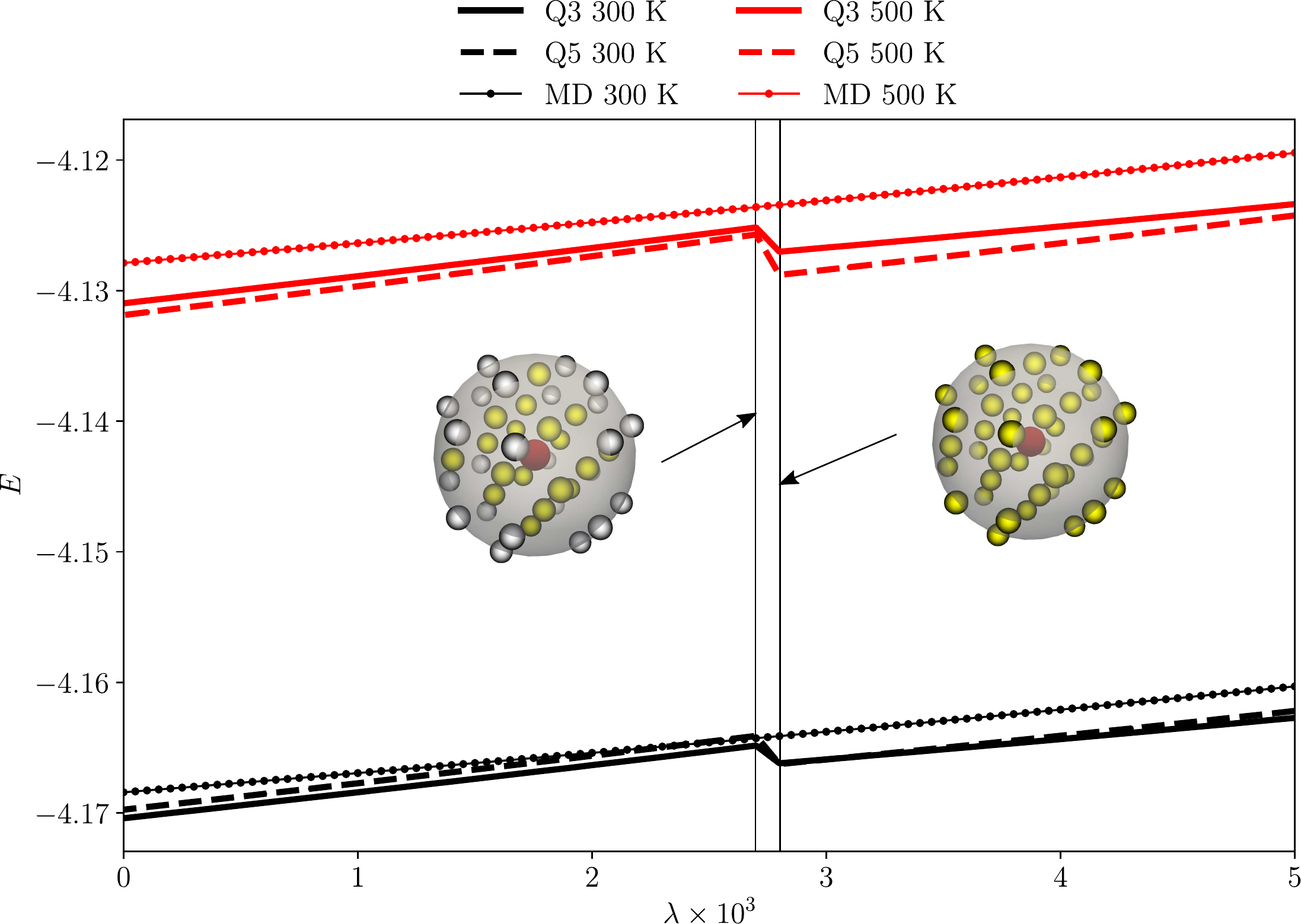}
\put(-240,100){$(a)$}
\put(-95,100){$(b)$}
\caption{Relaxed internal energy per atom in bulk iron in the FCC phase, using the \citet{meyer1998martensite} potential, as a function of the hydrostatic compressive strain $\lambda$, obtained from the GPP framework with third- and fifth-order quadrature rules, and in comparison with MD values. Black and red colors correspond to 300~K and 500~K, respectively. Insets (a) and (b) illustrate the stages right before and after the third nearest neighbors are considered as neighbors of the red-colored central atom.  
% Schematic showing the mean positions of nearest and second-nearest neighbors in an FCC lattice along with the radial cut-off of the central atom. The instantaneous positions of the bottom right second-nearest neighbor according to the third-order quadrature rule are shown in red. \PG{Units in y-label missing. Also, consider annotating within the plot at different $\lambda$ values.}  \DK{Should we maybe change the figure on the right into a cartoon of snapshots at different temperatures (making the atoms smaller and showing several snapshots) to highlight the problem?} 
}\label{fig:Fe_MeyerEntel_cutoff}
\end{figure}

When undergoing large deformations, atomic coordination numbers (i.e., the numbers of interacting neighbors) may change. Here, we show that the effect of atoms entering and leaving the cutoff radius of an atom in a statistically averaged sense is captured poorly by using quadrature sampling points for neighbors identified using the respective mean positions. In an MD simulation, this issue is tackled by updating the \textit{Verlet neighbor list} after a specified number of time steps, which is significantly less than the number of time steps over which averages are computed. Therefore, some atoms which are at mean distances larger than the potential's cut-off radius also contribute to each other's energy in a time-averaged sense. In the GPP and \textit{max-ent} approaches, this effect can be described by considering the Gaussian sphere of influence of every atom, which is centered around its mean position $\bar{\bfq}$ and has a width determined by its position variance $\Sigma$. As shown in the inset of Fig.~\ref{fig:Fe_MeyerEntel_cutoff}(a), 24 third-nearest neighbor atoms, whose mean positions fall outside the cut-off radius of the central atom (colored red) but which have a non-zero intersection of their Gaussian spheres of influence, should contribute to the central atom's energy with some non-zero probability. However, if the neighbors are identified only based on the mean positions of the atoms, the contribution of such neighbors are not accounted for in the calculation of phase averages. 

%Consequently, in a typical deformation scenario (hydrostatic compression considered in Fig.~\ref{fig:Fe_MeyerEntel_cutoff}), such neighbors abruptly enter the neighbor list of the atoms, thus causing spurious jumps in the phase-averaged quantities (internal energy per atom shown).

The effect becomes visible when computing the phase-averaged Hamiltonian of FCC iron at 300~K and 500~K as a function of an applied hydrostatic compressive strain. Interatomic interactions are modeled using the potential by \citet{meyer1998martensite}, which has two minima corresponding to the FCC and BCC phases of iron. The 0~K lattice spacing for the FCC phase is $3.61~$\AA. The third-nearest neighbors of an atom in a bulk environment lie right outside the cut-off radius at this spacing, and their effect is ignored as explained above. The quadrature rules sample the effect of only 12 nearest and 6 second-nearest neighbors. As the crystal is compressed, 24 third-nearest neighbors are identified to move inside the cut-off radius abruptly, at an approximate strain value of $2.7\times 10^{-3}$. As a consequence, a sudden drop in the phase-averaged energy is observed at this strain level in Fig.~\ref{fig:Fe_MeyerEntel_cutoff}. For reference MD simulations, we again use a fully periodic simulation cell of $10\times 10\times 10$ unit cells initialized in the FCC phase, from which we obtain time-averaged energies using the \textit{NVT} ensemble. Approximately 40 neighbors per atom are recorded for all strain values and, hence, MD values do not show the jump that is present in the GPP calculations.

Theoretically, each atom in the ensemble has a probability density for its instantaneous position, which asymptotically tends to zero only at infinite distances. Hence, all atoms have some (although infinitesimal for faraway atoms) effect on a certain atom's energy, and the average potential energy can be most accurately defined as
\begin{equation}
 \langle V_i(\bfq) \rangle = \int_{\Gamma} V_i(\bfq) \prod^{N}_{j=1} \text{exp}\left[ -\frac{ \vert \bfq_{j} - \bar{\bfq}_{j} \vert^2}{2 \Sigma_{j} }  \right] \text{d}\bfq_{j}  ,
 \label{eq: AllNeighbours}
\end{equation}
where the product now extends over all $N$ atoms in the ensemble. To properly approximate the average potential energy, confidence intervals can be defined, which impose a tolerance on the approximation. For example, we may define confidence intervals as 
\begin{equation}
    \vert \Bar{\bfq}_i - \Bar{\bfq}_j \vert \geq r_c + g(\Sigma_i,\Sigma_j) \quad \text{s.t.} \quad P(\vert \bfq_i - \bfq_j \vert \leq r_c ) = \epsilon,
\end{equation}
where $r_c$ is the cut-off radius of the potential, $g(\Sigma_i,\Sigma_j)$ is some function of the position variance of atom $i$ and its neighbor $j$, and $\epsilon>0$ is a small value. Physically, the confidence interval of $\epsilon$ implies that those atoms for which the probability of falling within $r_\mathrm{cut}$ of atom $i$ is less than or equal to $\epsilon$ are not considered in $n_N$. For sufficiently small $\epsilon$, the contribution of atoms which are close to the boundaries of a sphere of radius $r_\mathrm{cut}$ are accounted for in the phase average. Hence, the cut-off distance for mean positions (and consequently the coordination number of atom $i$) becomes a function of the position variances of the atoms and the confidence interval. This leads to the approximation
\begin{equation}
    \langle V_i(\bfq) \rangle = \int_{\Gamma} V_i(\bfq) \prod^{N}_{j=1} \text{exp}\left[ -\frac{ \vert \bfq_{j} - \bar{\bfq}_{j} \vert^2}{2 \Sigma_{j} }  \right] \text{d}\bfq_{j} \approx \int_{\Gamma} V_i(\bfq) \prod^{n_N(\Bar{\bfq}_i,\Bar{\bfq}_j, \Sigma_i, \Sigma_j, \epsilon )}_{j=1} \text{exp}\left[ -\frac{ \vert \bfq_{j} - \bar{\bfq}_{j} \vert^2}{2 \Sigma_{j} }  \right] \text{d}\bfq_{j},
\end{equation}
where $n_N(\Bar{\bfq}_i,\Bar{\bfq}_j, \Sigma_i, \Sigma_j, \epsilon )$ denotes the number of neighbors as a function of the mean atomic positions, position variances, and probability value $\epsilon$ chosen. By contrast, if only the cut-off of the interatomic potential (which is valid for $0~$K) and the mean positions of the atoms are used to define $n_N$, this eliminates the dependence of the coordination number on the atoms' position variances. Thus, results from Eq.~\ref{eq: AllNeighbours} are not the same as those evaluated with $n_N(\bar{\bfq}_i, \bar{\bfq}_j)$, i.e.,
\begin{equation}
    \langle V_i(\bfq) \rangle \approx \idotsint_{\Gamma} V_i(\bfq) \prod^{n_N(\Bar{\bfq}_i,\Bar{\bfq}_j, \Sigma_i, \Sigma_j, \epsilon )}_{j=1} \text{exp}\left[ -\frac{ \vert \bfq_{j} - \bar{\bfq}_{j} \vert^2}{2 \Sigma_{j} }  \right] \text{d}\bfq_{j} \neq \idotsint_{\Gamma} V_i(\bfq) \prod^{n_N(\bar{\bfq}_i, \bar{\bfq}_j)}_{j=1} \text{exp}\left[ -\frac{ \vert \bfq_{j} - \bar{\bfq}_{j} \vert^2}{2 \Sigma_{j} }  \right] \text{d}\bfq_{j}.
\end{equation}

\subsection{Insufficient accuracy} 
\label{Accuracy issue}

\begin{figure}[!tb]
\centering
\includegraphics[width = \textwidth]{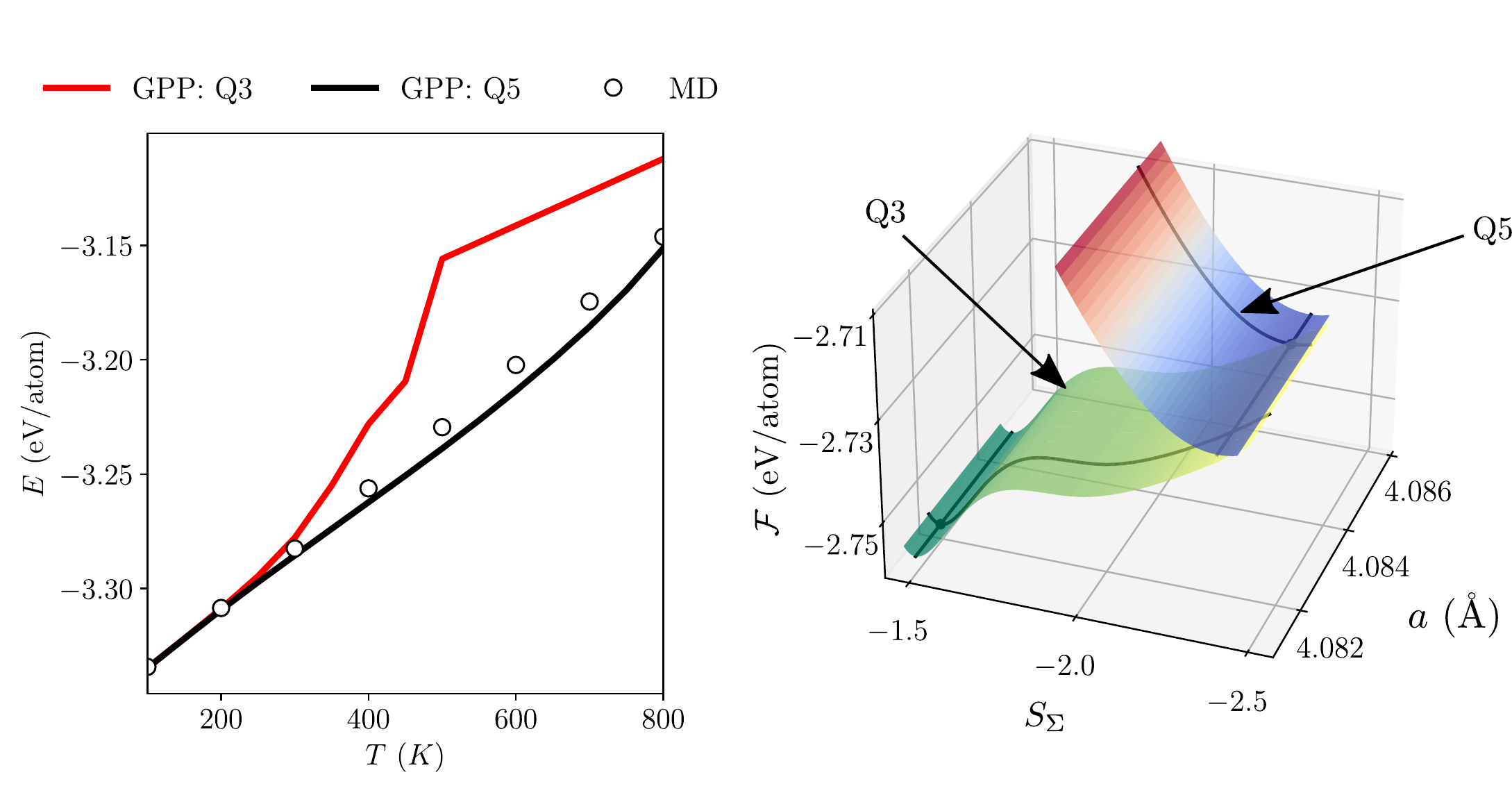}
\put(-460,10){$(a)$}
\put(-230,10){$(b)$}
\caption{(a) Relaxed average internal energy for bulk aluminium, calculated using the potential of \citet{mishin1999interatomic}, as a function of temperature obtained from the GPP framework with third- and fifth-order quadrature rules, and in comparison with MD values. (b) Helmholtz free energy landscape as a function of the FCC lattice spacing and positional entropy for bulk aluminium at 500~K with third- and fifth-order quadrature rules.
 }\label{fig:Al_mishin_inaccuracy}
\end{figure}

To demonstrate the inaccuracy incurred by the phase space quadrature rule, we show a particular example, where the third-order quadrature rule fails to approximate the phase space integral with sufficient accuracy, leading to spurious results. Fig.~\ref{fig:Al_mishin_inaccuracy} shows the thermal expansion of bulk aluminium, using the interatomic potential developed by \citet{mishin1999interatomic}. Average Hamiltonian values obtained from the GPP framework using third- and fifth-order quadrature rules are shown as a function of temperature from 100~K to 800~K, along-with MD values obtained using an \textit{NPH} ensemble with a Langevin thermostat. Third-order quadrature calculations deviate significantly from the MD values above 300~K, while the fifth-order quadrature results continue to agree well. Fig.~\ref{fig:Al_mishin_inaccuracy} shows the Helmholtz free-energy landscape as a function of the lattice spacing and the positional entropy  $(S_{\Sigma} = \ln{\Sigma}/2 )$ \cite{gupta2021nonequilibrium} at 500~K for the third- and fifth-order quadrature rules. As shown, the third-order energy landscape even displays a spurious minimum, which is a consequence solely of the inaccurate energy sampling by the Gaussian quadrature rule in phase space. 

\section{Machine learning-based quadrature surrogate models} \label{GNN architecture}
Motivated by the above shortcomings of Gaussian quadrature rules, we explore an alternative strategy in machine learning-based surrogate models. The universal approximation theorem for neural networks proves that multi-layer perceptrons (MLPs) with at least a single hidden layer and arbitrary width are universal function approximators and can be used to approximate any nonlinear continuous functional (such as the sought phase space integration of smooth interatomic potentials, mapping from a function space to the real numbers) under mild conditions \cite{Chen1993,Lu2021}. In our case, the challenge is to approximate the average potential energy $\langle V \rangle$ along with its partial derivatives with respect to mean atomic positions $(\bar{\bfq})$ and position variances $(\Sigma)$, which
represent the physical forces $\langle F \rangle$ and thermal forces $\langle F \cdot (\bfq - \bar{\bfq}) \rangle$, respectively, required for simulations. Given an interatomic potential $V$ as a function of the instantaneous atomic positions $\bfq$, we seek to train a neural network to learn the phase-averaged potential energy landscape $\langle V \rangle(\bar\bfq,\bfSigma)$ as a function of the mean atomic positions $\bar\bfq$ and the position variances $\bfSigma$ along with the multi-dimensional partial derivatives. To this end, we use MLPs as approximators to the sought phase space averages, effectively replacing expensive ‘\textit{online}’ quadrature rule computations during the simulations by a simple forward evaluation of an ‘\textit{offline}’ trained neural network.

Any sufficiently large MLP can approximate the sought mapping from mean atomic positions and position variances to the phase space average of the potential energy and the associated forces to any degree of accuracy if given enough training data and an optimizer that can identify the optimal parametrization of the network. However, experience has shown that under the more realistic assumption of limited resources (e.g., data and computing time) and typically considered gradient-based optimizers, inducing an \textit{inductive bias} into the network architecture can drastically improve the performance of deep learning architectures \cite{LeCun2015}. Inductive bias refers to specific assumptions on the structure of the data which can be incorporated into the architecture of the chosen deep-learning framework---the classical example being the translation-equivariant convolutional layers in a convolutional neural network, which are well-suited for image classification problems. In the latter, features of interest are typically not associated with a specific (absolute) pixel location \cite{Cohen2016}. Similarly, in atomistic scenarios, quantities of interest are not associated with individual atomic positions, rotations of atomic clusters, and ordering of input variables. In earlier attempts for ML-based interatomic potentials, this was achieved by mapping the information of atomic positions to a set of structural parameters/descriptors that encode the local environment of an atom \cite{onat2020sensitivity}. High-dimensional regression models are then used to map the structural descriptors onto the potential energy landscape \cite{mishin2021machine}. 

\subsection{Equivariant graph neural networks}

Convolutional neural networks can be understood as a specific instance of a GNN that operates on a regular 2D grid. GNNs in their general form can operate on an arbitrary graph structure, represented by a set of nodes and edges. This is often a useful abstract representation of data structures found in physical systems---for further details of GNNs and their general functionality we refer to \cite{Zhou2020}. We here exploit the analogy with atomic ensembles, in which each atom is represented as a node and the pairwise interactions between atoms are the corresponding edges between nodes. Unlike standard MLPs, GNNs possess intrinsic biases well-suited towards such graph structures. Most notably, GNNs typically contain a permutation-invariant aggregation operator over adjacent node features, and thus any arbitrary permutation of the graph nodes (i.e., input order of the atoms) is invariant to the prediction of the network, allowing it to focus on the structure of the graph. In the context of molecular chemistry, the overall energy of a molecule should also be invariant to translations, rotations, and reflections of the overall molecule. This implies that the GNN operator should be invariant to E(3) group isometries. Harmonic Networks \cite{worrall2017harmonic}, SchNet \cite{schutt2017quantum} and Tensor field networks \cite{thomas2018tensor} are examples of GNN networks that incorporate translational and rotational invariance. This can most easily be achieved by considering relative interatomic distances as input to the GNN \cite{Schutt2017}, while invariance with respect to reflections can be ensured by a simple selection rule \cite{batzner20223}.

Though such networks respect the physically motivated invariance of such operations with respect to scalar quantities such as the energy, research has shown that the possibility of learning internal geometric representations that behave \textit{equi}variant to rotations and reflections can be beneficial for training and generalization. This may furthermore be desirable for predicted \textit{tensorial} quantities such as force vectors (which may be derived from the predicted energy), as those should rotate consistently with the atomic ensemble. This can be achieved, among others, by constructing a convolutional filter as a product of (fixed) $SO(3)$-equivariant spherical harmonics and a rotationally invariant (learnable) radial function, which thus inherits $SO(3)$-equivariance (see \cite{Thomas2018} for further details). Motivated by their reported success in developing interatomic potentials \cite{batzner20223}, we consider such architectures and adjust them to our scenario.

\subsection{Application to the GPP formalism}

We consider the recent NequIP framework \cite{batzner20223} as the base for our sought GNN surrogate model. The original framework was designed to build interatomic potentials for MD. We here extend the general approach but adapt it to approximate the phase space integral of the energy, $\langle V\rangle$, and the associated forces for a given interatomic potential $V$. The thus-obtained quadrature surrogate is included in our GPP framework---thereby shifting the computational complexity of the quadrature computation to an offline training procedure, while inference during simulations is drastically reduced. In addition to the mean atomic positions $\bar\bfq$, we consider the positional variances $\bfSigma$ of atoms as an additional continuous (scalar) node feature, which is passed to our model  with $\bar{\bfq}$. 

To compute the training data for the phase space average (i.e., average energies and corresponding forces), we rely on a highly accurate Monte-Carlo integration scheme with the Metropolis algorithm \cite{hastings1970monte}. Accordingly, we approximate Eq.~\eqref{Quadrature} as
\begin{equation}
    \langle V_i \rangle \approx 
\frac{1}{N_{MP}} \sum_{p=1}^{N_{MP}} V_i(\bfq_p),
\end{equation}
where $N_{MP}$ is the number of uncorrelated sampling points. A sequence of these sampling points is generated with the GPP probability density of every atom being the acceptance probability density, as shown in~\eqref{independent GPP}. This ensures that our Monte-Carlo scheme executes an importance sampling over the phase space of atomic positions weighted by the relevant probability density \cite{frenkel2001understanding}. To ensure uncorrelated samples, we discard $3n$ samples between every two sampling points, where $n$ is the number of atoms in the cluster. 

In finding the optimal training protocol, we observed that training data generated from amorphously displaced atoms resulted in accelerated loss convergence compared to training data generated from uniform deformations of crystalline structures. Therefore, we proceed as follows. Given an interatomic potential, we obtain the relaxed mean lattice spacing and relaxed position variance of an atom in the bulk over a range of temperatures from 100~K to 500~K in steps of 50~K, using the third-order Gaussian quadrature. This serves as a rough estimate of the relaxed lattice spacing at different temperatures, while the actual phase-averaged energy and forces for training and validation data are obtained from the Monte-Carlo sampling scheme. At every temperature, independent atomic clusters are generated with $\bar{\bfq}$ and $S_\Sigma$ defined using the corresponding relaxed lattice parameter and bulk position variances. A Gaussian noise of $\calN(\bar{\bfq}, r_c / 50 )$ is applied to the mean positions, and $\calN(S_{\Sigma}, 0.25 )$ is applied to the positional entropies of all atoms in the cluster to generate different data points. Metropolis based Monte-Carlo sampling is then applied to clusters generated in this fashion. We use the actual position variances $\Sigma = \exp(2S_\Sigma)$ instead of the positional entropies $S_\Sigma$ as the node features and additionally normalize them by their root mean square (RMS) values, as this was observed to increase training stability. Note that as in the original NequIP framework \cite{batzner20223}, the predicted energies are scaled to zero mean and unit variance. Physical and thermal forces are computed from the learned average potential energy landscape $(\hat{E})$ as
\begin{align}
&\langle \itbf{F}_{ij} \rangle = -  \frac{ \partial\hat{E}_i} {\partial \bar{\bfq}_j }, \nonumber \\
& \langle F_{ij}^\text{th} \rangle =   \frac{\langle \itbf{F}_{ij} \cdot (\bfq_j - \bar{\bfq}_j) \rangle}{2\Sigma_j} = 
  \frac{ \partial\hat{E}_i} {\partial \Sigma_j }.
\end{align}
where $\itbf{F}_{ij}$ and $F_{ij}^{\text{th}}$ denote, respectively, the force and thermal force on atom $j$ due to atom $i$, and $V_i$ is the potential energy of atom $i$.

In the GPP framework, we consider independent atomic clusters to solve Eqs.~\eqref{EOM quasistatic} at every atomic site by calculating the effect of all interacting neighbors of that particular site. Therefore, we treat the entire atomic ensemble as a collection of $N$ independent graphs---each graph consisting of an atom $i$ as the central atom, which is connected to its neighbors, thus corresponding to a star graph. As we are only interested in the phase space average energy of the central atom, we collect the node feature of the central atom after four message passings (convolutional layers) to predict a single scalar quantity, which is the phase-averaged potential energy $\langle V_i\rangle$ of the central atom $i$. First-order derivatives are computed using a backward pass through the graph. We use the combined mean-squared-error (MSE) of the energy, physical forces, and thermal forces as the loss function without additional (relative) scaling of the loss terms. Consequently, the loss function can be written as
\begin{equation}
\label{loss function}
    \calL_i = \left ( \langle V \rangle_i - \hat{E}_i \right) ^2 + \frac{1}{3n_i} \sum_{j=1}^{n_i} \left\lVert \langle \bfF_{ij} \rangle + \frac{ \partial \hat{E}_i } {\partial \bar{\bfq}_{j}} \right\rVert^2 + \frac{1}{n_i} \sum_{j=1}^{n_i} \left (  \frac{\langle  \itbf{F}_{ij} \cdot (\bfq_j - \bar{\bfq}_j ) \rangle}{ 2 \Sigma_j } -   \frac{ \partial \hat{E_i}} {\partial \Sigma_j } \right)^2
\end{equation}
where $\hat{E}$ represents the predicted potential energy landscape, $n_i$ is the number of atoms, and $\calL_i$ is the loss for the $i^{th}$ data point. The total loss is computed as the average over all data points. 
This constitutes the training setup for the E(3)-equivariant GNN, which is used within the GPP framework for time-coarsened atomistic simulations---replacing the costly and inaccurate Gaussian quadrature in phase space to calculate forces and thermal forces on all atoms in the ensemble. A detailed summary of the GPP implementation was provided by \citet{gupta2021nonequilibrium}.
% A schematic of the GPP solution procedure using a GNN model is shown in Fig.~\ref{fig:flowchart}.  

% \begin{figure}[!hb]
% \centering
% \input{flowchart.tex}
% \caption{Solution procedure of the GPP scheme using the GNN. $L_1$ denotes the $L_1$-norm of equation \eqref{EOM quasistatic} while $\delta$ is the tolerance value chosen for convergence.}
% \label{fig:flowchart}
% \end{figure}
\begin{table}[!b]
\centering
\begin{tabular}{| c | c | c |  c |  c | c|  m{0.8cm} | m{0.8cm} | m{0.8cm} | m{0.8cm} |  m{0.8cm}| m{0.8cm} |  }  \hline
& \multirow{3}{1.2cm}{Dataset size} &  \multirow{3}{1.2cm}{Tensor rank $(l)$} &  \multirow{3}{1.2cm}{No. of node features} & \multirow{3}{1.0cm}{Radial cutoff (\AA) } & \multirow{3}{1.0cm}{Batch size } &  \multicolumn{2}{c|}{ \multirow{2}{1.8cm} {  $\langle V \rangle$ (meV) }} &  \multicolumn{2}{c|}{  \multirow{2}{2.0cm} {  $\langle \itbf{F} \rangle$(meV/\AA) }} & \multicolumn{2}{c|}{  \multirow{2}{2.0cm} {  $\langle F^\text{th} \rangle$(meV/\AA$^2$) }} \\

& & & & & & \multicolumn{2}{c|}{}  & \multicolumn{2}{c|}{}  & \multicolumn{2}{c|}{}
\\ \cline{7-12}

& & & & & & MAE & RMSE & MAE &RMSE&MAE&RMSE \\ \hline 
Cu & $5\times 10^5$ & $1$ & $64$ & $4.5$ & $32$ & $0.613$ & $0.749$  & $0.110$ & $0.237$ &  $0.764$ & $1.46$ \\ \hline
Fe & $1\times 10^6$ & $1$ & $64$ & $4.4$ & $32$ & $0.269$ & $0.378$ & $0.189$ & $0.360$ & $0.362$  & $0.551$ \\ \hline
Al & $2.5\times 10^5$ & $1$ & $64$ & $6.4$ & $32$ & $0.313$ & $0.396$ & $0.0734$ &  $0.162$ & $0.232$ & $0.807$ \\ \hline
\end{tabular}
\caption{Maximum absolute error (MAE) and root mean square error (RMSE) of the potential energy, force, and thermal force along with training hyperparameters for phase-averaged Cu using the EFS potential~\cite{dai2006extended}, Fe based on the Meyer-Entel potential~\cite{meyer1998martensite}, and Al using the Mishin potential~\cite{mishin1999interatomic}. Errors are computed over the validation set, which makes up 10\% of each dataset.}
\label{Training table}
\end{table}
Table~\ref{Training table} summarizes the dataset size, training hyperparameters, and validation metrics for the computed energy, forces, and thermal forces of the three models we will be comparing in Section~\ref{results}. Apart from those mentioned there, all hyperparameters (including the number of convolutional layers, radial network basis, learning rate, activation functions, and optimizer choice) were kept to be the same as in \citet{batzner20223}.

\section{Results} \label{results}
In this section, we present simulation results obtained from the GPP methodology combined with both Gaussian quadrature and GNN based phase-space averaging techniques for time-coarsened atomistics. These results illustrate the efficiency and accuracy of the GNN model as compared to third- and fifth-order Gaussian quadrature rules (and compared to reference data obtained from MD). A key advantage of the GPP technique as opposed to MD is its ability to compute quasistatic properties from atomistics without the need for costly MD simulations and subsequent time averaging. Instead, a single quasistatic relaxation step is sufficient (relaxing both atomic forces and thermal forces simultaneously) to arrive at finite-temperature average quantities such as internal energies. This is demonstrated in our examples. In Section~\ref{results 1}, we start with the simple example of simulating the thermal expansion of bulk copper, using the extended Finnis-Sinclair \cite{dai2006extended} potential. We compare the lattice parameter and average Hamiltonian as functions of temperature with MD data. As a second example, we show the Helmholtz free energy variation along the Bain path for the temperature-driven martensitic transformation in bulk iron, using the Meyer-Entel \cite{meyer1998martensite} potential in Section~\ref{results 2}. For reference, the obtained data will be compared to MD results obtained from thermodynamic integration \cite{freitas2016nonequilibrium}. Finally, in Section~\ref{results 3} we compare the compare the free energy of samples with grain boundaries in aluminum and FCC iron. We demonstrate that the GNN-based phase space quadrature within the GPP simulation framework proves to be a robust and accurate technique, alleviating the issues discussed in Section~\ref{List of problems}.

\subsection{Thermal expansion of copper} 
\label{results 1}
\begin{figure}[!t]
\centering
\includegraphics[width = \textwidth]{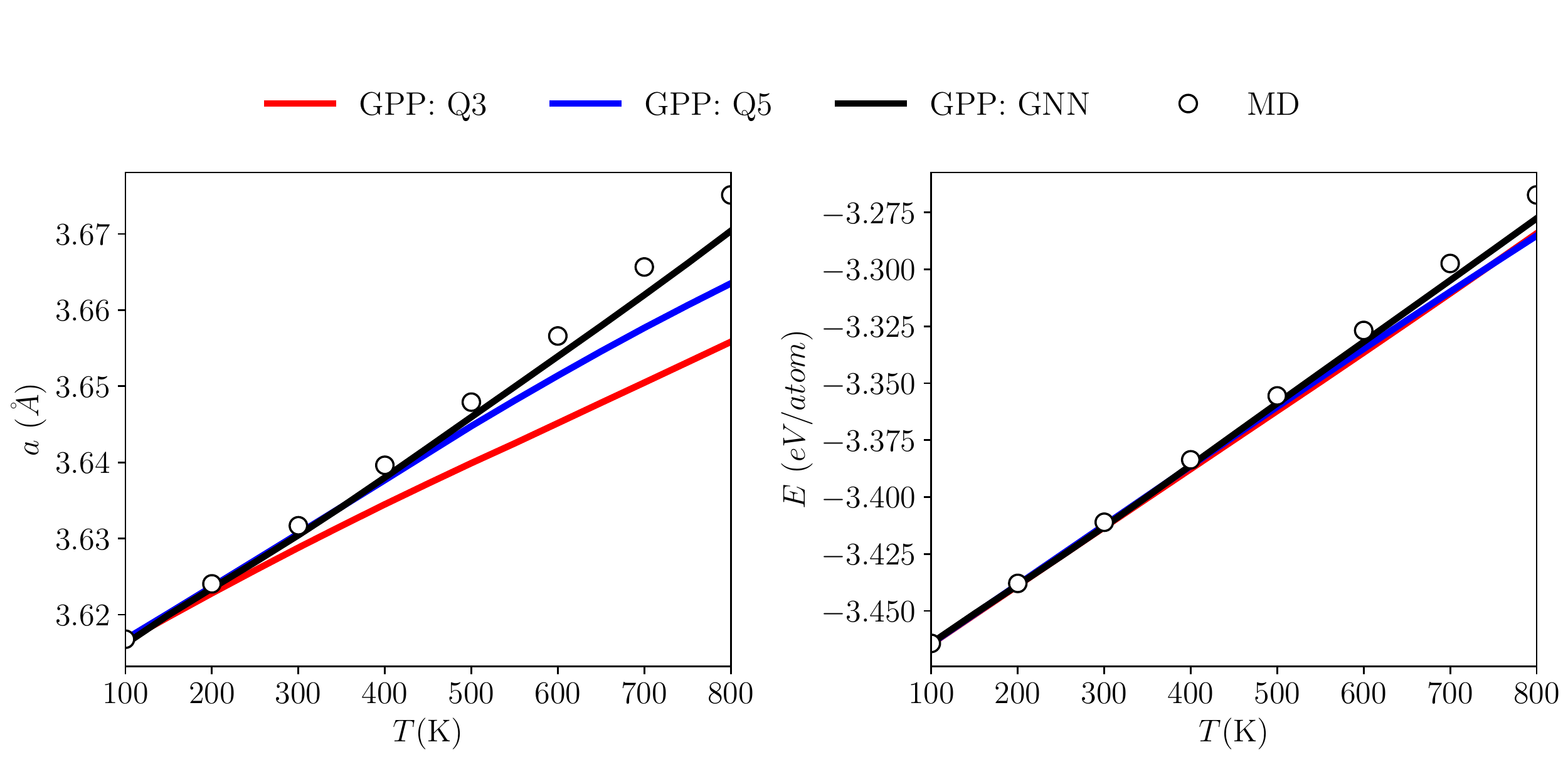}
\caption{Lattice parameter $a$ and average Hamiltonian $E$ as a function of temperature for copper modeled by the EFS potential \cite{dai2006extended}, comparing results obtained from the GPP framework with third- and fifth-order Gaussian quadrature and from the proposed GNN-based approximator as well as from MD.}\label{fig:copper thermal expansion}
\end{figure}

In Fig.~\ref{fig:copper thermal expansion} we show the lattice parameter $a$ and the average Hamiltonian/internal energy $E$ per atom for bulk copper from 100~K to 800~K. The shown data were obtained from the GPP framework with various quadrature types as well as from MD. For MD simulations, we use a fully periodic $10\times 10\times 10$ unit cell box and relax the box dimensions over the duration of 10~ns ($10^7$ time steps at a step size of 1~fs), using the $NPH$ ensemble with the Langevin thermostat in LAMMPS. By comparison, GPP simulations require only a single cluster (an atom with its interacting neighbors), for which relaxation finds the equilibrium bulk configuration. As seen in Fig.~\ref{fig:copper thermal expansion}, the GNN-based GPP values for the internal energy stay within 0.32\%, and those for the lattice parameter stay within 0.13\% of the reference MD values up to temperatures as high as 800~K. Deviations between GPP and MD are observed to increase with temperature in general due to missing interatomic correlations in the independent GPP ansatz \eqref{independent GPP} and limited quadrature sampling accuracy. This is expected and inherent in all comparable time-coarsened atomistic frameworks. Using GNN models trained with high-quality Monte-Carlo data results in deviations that arise solely from the missing information of interatomic correlations in the statistical mechanics-based ansatz for the probability distribution, but not from numerical quadrature. Data obtained from the same GPP framework with third- and fifth-order Gaussian quadrature display large errors at significantly lower temperatures.

\subsection{Martensitic phase transition in iron} 
\label{results 2}

The martensitic BCC-to-FCC phase transition in iron is one of the most well known temperature-driven transitions in structural metals---which we use in this section to demonstrate that the inobjectivity and discontinuity issues of quadrature rules (discussed in Section~\ref{List of problems}) make the GPP framework inapplicable for scenarios involving phase transitions. The transition from the BCC ($\alpha$-iron/ferrite) to the FCC ($\gamma$-iron/austenite) phase has been experimentally observed \cite{lee2012atomistic} to occur at a temperature of 1173~K. However, only few interatomic potentials capture this $\alpha\leftrightarrow \gamma$ transition. As confirmed by \citet{meiser2020alpha}, the Meyer-Entel \cite{meyer1998martensite} potential is the only EAM potential that shows the existence of two energy minima corresponding to the ferrite and austenite phases of iron. Others include the MEAM potential by \citet{lee2012atomistic} and a bond-order potential by \citet{muller2007analytic}. In the following, we use the EAM potential of \citet{meyer1998martensite}.

\begin{figure}[h!]
\centering
\includegraphics[width = \textwidth]{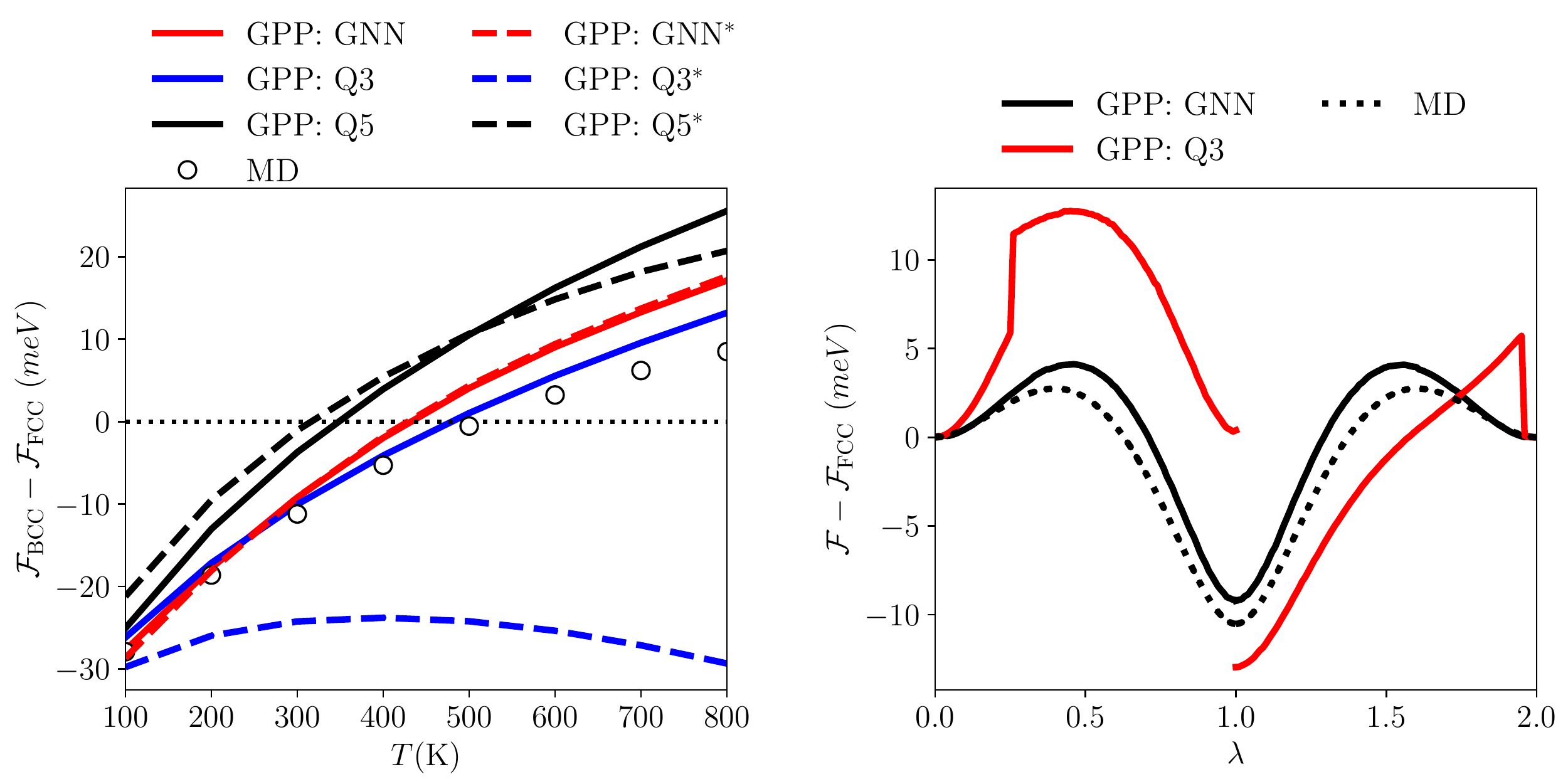}
\put(-460,10){$(a)$}
\put(-220,10){$(b)$}
\caption{(a) Helmholtz free energy difference betweeen the BCC and FCC phases of iron vs.\ temperature, as obtained from the GNN-based model, third- and fifth-order Gaussian quadrature, and from MD. Solid lines indicate computations using the exact cutoff radius of the Meyer-Entel potential, while dashed lines show computations with a slightly larger cutoff radius.  (b) Helmholtz free energy along the Bain path (vs.\ the compressive strain parameter $\lambda$) at 300~K, shifted by the Helmholtz free energy of the FCC phase.}\label{fig:iron phase transition}
\end{figure}

Fig.~\ref{fig:iron phase transition}(a) shows the Helmholtz free energy difference between the BCC and FCC phases per atom for bulk iron as a function of temperature. This data was obtained from relaxing separately a single atom cluster in the BCC and FCC phases (thus mimicking an infinite crystal) from 100~K to 800~K, using the GPP framework. For reference, we compute the same free energies via MD, using a $10\times 10\times 10$ simulation cell in LAMMPS with the \textit{NPH} ensemble and a Langevin thermostat. It is important to note that we use a cubic box in LAMMPS, so that the lattice structure does not by itself show the $\alpha \rightarrow \gamma$ transition with increasing temperature. (Such a transition can be observed if a triclinic box is used instead \cite{meyer1998martensite}.) The temperature at which the shown energy difference changes sign is the critical temperature of the phase transition. As seen in Fig.~\ref{fig:iron phase transition}(a), the third-order quadrature rule gives values close to MD values, when the exact cutoff radius of the Meyer-Entel potential is used. However, it severely miscalculates the energy difference, when a slightly larger cutoff is used (see the curves labeled Q3$^*$ and Q5$^*$). Moreover, it shows the BCC phase to be stable across the whole temperature range. This happens due to erroneously computed contributions from $24$ third-nearest neighbors in the FCC phase, which lie marginally outside the true potential cutoff. The fifth-order quadrature does predict the transition but deviates considerably from the MD values. Our GNN model yields energy difference values significantly closer to those obtained from MD and also correctly captures the contribution of atoms lying slightly outside the true potential cutoff. 

We proceed to compute the free energy along the Bain path \cite{bain1924nature} at 300~K for an FCC to BCC transition and vice versa. Although not close to the actual reaction pathway for the martensitic phase transition, we consider the Bain path, since it is a simple triaxial strain parametrization that clearly allows us to demonstrate the limitations of Gaussian quadrature rules. A volume-preserving Bain transformation is achieved by expanding an FCC lattice along the $[100]$ and $[010]$ directions by approximately $12\%$ and contracting along the $[001]$ direction by approximately $21\%$ \cite{sandoval2009bain}. As the volume of the atomic unit cell in the two phases is not the same, the actual strain values imposed to transform an FCC lattice at zero pressure to a BCC lattice at zero pressure deviate slightly from the values mentioned above. The values are identified for 300~K using the lattice vectors obtained from the individual thermal expansion of the two phases. For the third-order quadrature and GNN calculations, the $\gamma\to\alpha$ transition is characterized by
\be
    \boldsymbol{F}^{\mathrm{Q3}}_{ \gamma \rightarrow \alpha } = \begin{bmatrix} 1+0.1046 \lambda & 0 & 0 \\ 0 & 1+0.1046\lambda & 0 \\ 0 & 0 & 1-0.2173\lambda \end{bmatrix}, \quad  \boldsymbol{F}^{\mathrm{GNN}}_{ \gamma \rightarrow \alpha } = \begin{bmatrix} 1+0.1109\lambda & 0 & 0 \\ 0 & 1+0.1109\lambda & 0 \\ 0 & 0 & 1-0.2145\lambda \end{bmatrix}, 
\ee
while the reverse $\alpha\to\gamma$ transition follows
\be
    \boldsymbol{F}^{\mathrm{Q3}}_{ \alpha \rightarrow \gamma } = \begin{bmatrix} 1-0.0947 \lambda & 0 & 0 \\ 0 & 1-0.0947\lambda & 0 \\ 0 & 0 & 1+0.2775\lambda \end{bmatrix}, \quad   \boldsymbol{F}^{\mathrm{GNN}}_{ \alpha \rightarrow \gamma } = \begin{bmatrix} 1-0.0998\lambda & 0 & 0 \\ 0 & 1-0.0998\lambda & 0 \\ 0 & 0 & 1+0.2730\lambda \end{bmatrix}. 
\ee
Difference between GNN and Q3 stem from differences in the relaxed lattice spacing obtained from both approaches. For both the $\gamma \rightarrow \alpha$ and $\alpha \rightarrow \gamma$ transitions we compute the free energy at 100 increments of the strain parameter $\lambda$ along the Bain path.

For the GPP simulations, we initialize an atomic cluster in the relaxed FCC lattice spacing at 300~K and impose the displacement gradient to transform it into a zero-pressure relaxed BCC state in 101 uniform strain increments. The position vectors of all atoms are frozen at every straining step, imposing an affine deformation of the lattice, and a thermal relaxation is performed at each strain level by iteratively solving Eq.~\eqref{EOM2}. For MD simulations, we parallelly initialize 101 simulation boxes of $10\times 10\times 10$ unit cells and populate them with atoms generated according to the relaxed lattice vectors at the different strain increments. For every simulation box, we use the \textit{NVT} ensemble with a Langevin thermostat and the \textit{fix ti/spring} command in LAMMPS. While doing the Frenkel-Ladd path integration, the previously computed $\langle\vert\Delta \bfq\vert^2\rangle$ values are used to find the spring constants for the Einstein crystal approximation. Long equilibration times of $10$~ns are used to obtain accurate time averages, while the thermodynamic switching between the original system and the quasi-harmonic approximation is performed in $1$~ns. 
The analogous procedure is used to parameterize the (reverse) BCC to FCC transition in 101 uniformly spaced steps. 
The results are summarized in Fig.~\ref{fig:iron phase transition}(b), where $\lambda$ is the reaction coordinate (strain parameter) defined such that $\lambda=0$ to $1$ corresponds to the martensitic transition from FCC to BCC iron, and $\lambda=1$ to $2$ represents the reverse transition.  

The data obtained from third-order quadrature display striking discontinuities and jumps along the free energy path. Moreover, the free energy per atom of the martensite phase ($\lambda=1$) obtained after transforming from the FCC phase is different from an independently initialized lattice in the BCC phase. This is due to the orientational relationships $(001)_{fcc} \parallel (001)_{bcc}$ and $ [100]_{fcc} \parallel [110]_{bcc} $ for the Bain path, and the quadrature rule's lack of frame indifference with respect to $SO(3)$ rotations, as described in Section~\ref{Objectivity issue}. In addition, the transition from FCC to BCC and vice versa also shows discontinuities in the free energy. This stems from the fact that the number of neighbors with mean positions inside the cutoff radius changes from $18$ to $26$ at $\lambda=0.26$ and returns to $18$ at $\lambda=1.96$ during the reverse transition. Free energy data generated with the GNN model, on the other hand, are free of discontinuities and approximate well the MD data.

Aside from the level of inaccuracy, the existence of discontinuities in the free energy landscape renders the GPP framework with Gaussian quadrature not applicable for problems involving phase transitions, because finding energy barriers and minimum energy pathways for transitions via techniques such as the Nudged Elastic Band (NEB) \cite{henkelman2000improved} or the Dimer methods \cite{henkelman1999dimer} requires the existence of a smooth energy landscape to search for saddle points. The presented GNN model, by contrast, allows for the efficient use of the GPP framework for accurately approximating the transformation energy landscape without costly MD calculations, as illustrated here for the martensitic transformation in iron.

\subsection{Grain boundary energies} \label{results 3}

\begin{figure}[!h]
\centering
\includegraphics[width = \textwidth]{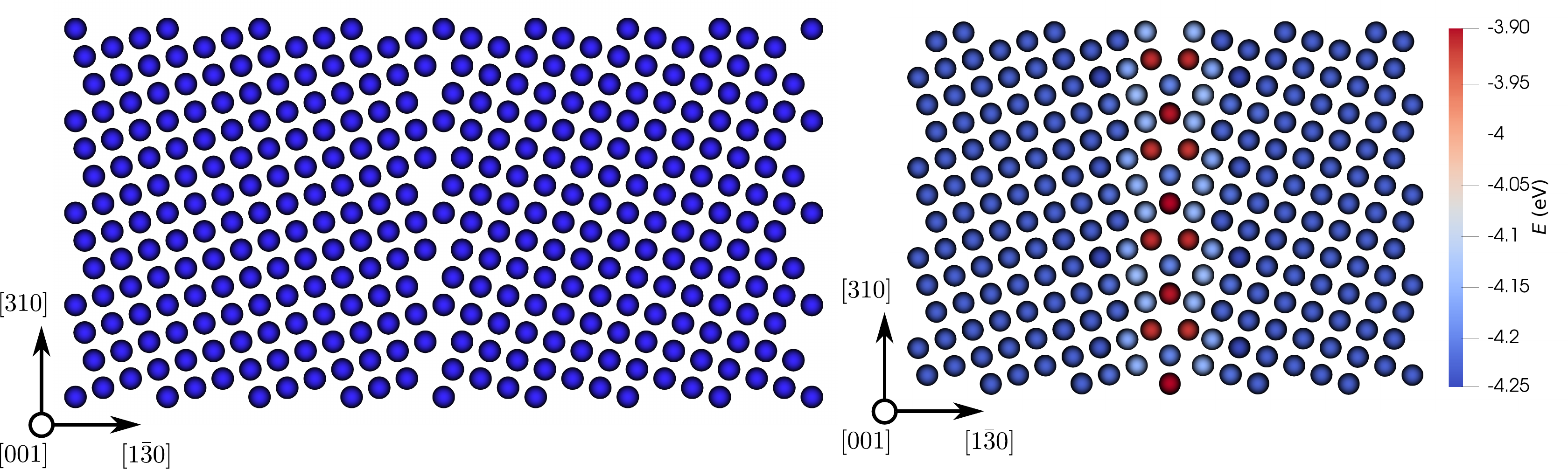}
\put(-460,-8){$(a)$}
\put(-210,-8){$(b)$}
\caption{(a) The $\Sigma5(310)[001]$ FCC grain boundary sample used for energy calculations. Periodic boundary conditions are used in all three dimensions, resulting in two grain boundaries: one in the center of the sample and another one at its boundary. (b) Sample with a relaxed FCC iron grain boundary at 300~K with atoms color-coded according to their average Hamiltonians.}
\label{fig:GB_samples}
\end{figure}

To illustrate the performance of the GNN-based GPP simulations, our last example computes the internal energy of FCC samples containing symmetric $\Sigma 5[310](001)$ grain boundaries. We perform these simulations for both aluminium, using the Mishin \cite{mishin1999interatomic} potential, and for iron in its FCC phase, using the Meyer-Entel \cite{meyer1998martensite} potential.
The sample is set up with periodic boundaries in all three dimensions, as shown in Fig.~\ref{fig:GB_samples}(a), resulting in two identical grain boundary planes: one in the center of the sample and the other one at its boundary. The lattice spacing used to generate the FCC crystals on both sides of the grain boundary is the relaxed bulk lattice spacing at the corresponding temperature. All of the samples have a total of 1280 atoms. For MD validation, we construct equivalent geometries in LAMMPS, with relaxed bulk lattice parameters obtained from prior MD runs. Samples in MD were relaxed for $10^7$ time steps with the \textit{NVE} ensemble and a Langevin thermostat.

\begin{figure}[!h]
\centering
\includegraphics[width = \textwidth]{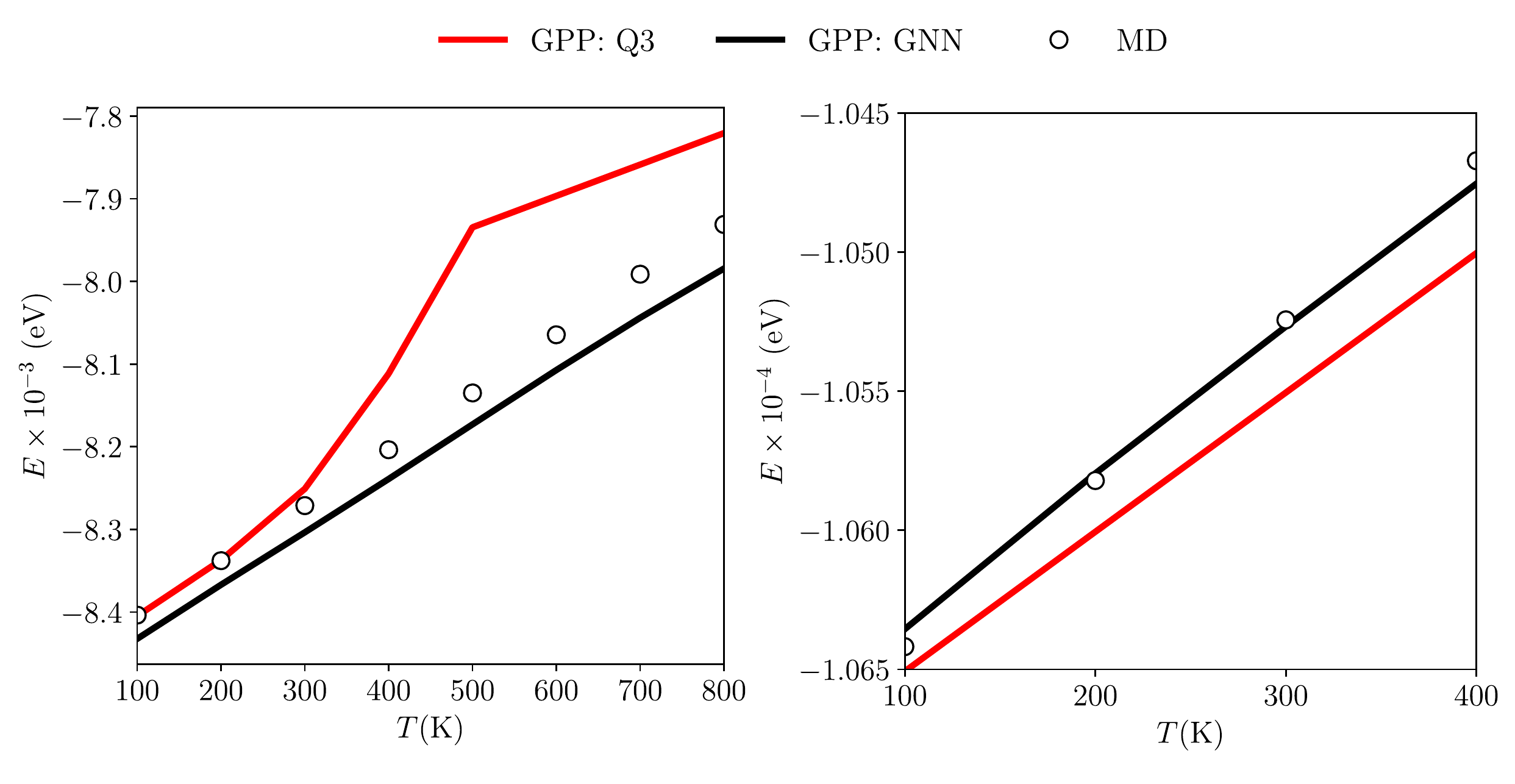}
\put(-460,8){$(a)$}
\put(-200,8){$(b)$}
\caption{(a) Relaxed internal energy of aluminum grain boundary samples with the Mishin potential \cite{mishin1999interatomic}, using third-order quadrature and GNN-based model and the comparison with MD data. (b) Relaxed internal energy of FCC iron grain boundary samples with the Meyer-Entel potential \cite{meyer1998martensite}, using the GNN-based model and the comparison with MD data.}
\label{fig:GB_energies}
\end{figure}

Fig.~\ref{fig:GB_energies}(a) shows the computed potential energy of the aluminum samples across a wide range of temperatures, as obtained from the GPP framework with both GNN-based phase space averaging and Gaussian quadrature as well as from MD for reference. The third-order quadrature produces a spurious jump in the internal energy at around 500~K, while the GNN-based model is robust and fairly accurate up to temperatures as high as 800~K. We note that the accuracy of simulations that use the GNN models can further be increased by training the model with a higher number of data points. As shown in Table~\ref{Training table}, significantly less training data points were used for the Mishin potential, which leads to the deviations from MD values in Fig.~\ref{fig:GB_energies}(a).

Knowing that the Meyer-Entel potential for iron was trained with a large number of samples, we also compute grain boundary energies for iron in its FCC phase to confirm the desired effect of using a larger number of data points. Fig.~\ref{fig:GB_energies}(b) shows the internal energies of iron grain boundary samples computed using the GNN-based model for and the reference MD values, both based on the Meyer-Entel potential. (A relaxed grain boundary sample at 300~K is shown in Fig.~\ref{fig:GB_samples}(b), where the atoms are color-coded with their averaged Hamiltonians.) The data in Fig.~\ref{fig:GB_energies}(b) indeed confirm that using a larger number of training points leads to more accurate predictions, as the agreement with MD is considerably improved as compared to Fig.~\ref{fig:GB_energies}(a) and the third-order quadrature results.

\section{Conclusion} 
\label{conclusions}

We have shown how high-dimensional phase space integrals, which appear ubiquitously in time-coarsened atomistic techniques such as DMD, max-ent and GPP, can be efficiently and accurately approximated by graph neural networks (GNN). We have demonstrated that the classically used alternative, viz.\ the use of Gaussian quadrature rules, comes with detrimental problems: (i)~such rules lack frame indifference, so that results depend on rigid-body crystal rotations; (ii)~challenges associated with the definition of atomic neighborhoods in the presence of statistical variations and cutoff radii; and (iii)~ a general lack of accuracy leading to fictitious energy minima. As a remedy for those issues, we have trained equivariant GNNs based on highly accurate Monte-Carlo integration data, which predict the phase-space-averaged energy of an atomic cluster and the associated atomic forces. By combining this data-driven approach with a Gaussian phase packet (GPP) formalism for time-coarsened atomistics, we train surrogate phase space integrators, which---for given atomic mean positions $\{\bar{\bfq}\}$ and displacement variances $\{\Sigma\}$--- predict the mean potential energy $\langle V \rangle$ and mean atomic forces. These, in turn, are used within the GPP framework (assuming interatomic independence and spherical Gaussian displacement variances) to simulate atomistic ensembles at finite temperature in a quasistatic fashion, thus bypassing costly MD simulations and the associated long-time averaging required to obtain converged phase space average data.

To validate and showcase the advantages of the trained GNN over the classically used quadrature rules, we have presented three benchmarks. First, simulating the thermal expansion of copper revealed superior accuracy of the GNN-based formulation compared to third- and fifth-order Gaussian quadrature rules over the entire range of temperatures studied (in comparison to MD). Second, we have studied the martensitic phase transition of pure iron, computing the free energy difference between the ferritic and austenitic phases as well as the energy landscape along the Bain path. Results showed that Gaussian quadrature is inapplicable in this scenario, while the GNN-based formulation provides a good match with MD data. Third, the general applicability of the GNN-based formulation within a GPP framework to simulate crystalline atomic ensembles with defects was illustrated by computing the energy of a symmetric tilt grain boundary. Again, the GNN data provided superior accuracy over Gaussian quadrature, as verified in comparison with MD.

In summary, we have presented an accurate and efficient data-driven method to compute high-dimensional phase space integrals, which overcomes the limitations of Gaussian quadrature. The latter is the de-facto standard technique used across time-coarsened atomistics as well as other statistical mechanics-based multiscale modeling techniques. Therefore, the presented approach offers applicability beyond our specific focus on GPP-based time-coarsened atomistics.
% shown that the results provided when replacing the quadrature rules with the GNN show very good agreement with the MD reference data in every scenario and avoid the problems that come with the quadrature while showing higher accuracy. Therefore, it has been demonstrated that these predictive models expand the applicability of the GPP formulation and equivalent frameworks, and have the potential to unlock and improve the accuracy \MS{(and efficiency?)} of other techniques beyond atomistics where high-dimensional ensemble integrals need to be repeatedly evaluated. \MS{possible extensions or future work}

% \appendix
% \section{My Appendix}
% Appendix sections are coded under \verb+\appendix+.

% \verb+\printcredits+ command is used after appendix sections to list 
% author credit taxonomy contribution roles tagged using \verb+\credit+ 
% in frontmatter.

\section*{Acknowledgements}
The support from the European Research Council (ERC) under the European Union’s Horizon 2020 research and innovation program (grant agreement no.~770754) is gratefully acknowledged.

\printcredits

\bibliographystyle{cas-model2-names}
\bibliography{bibliography}

\begin{thebibliography}{81}
\expandafter\ifx\csname natexlab\endcsname\relax\def\natexlab#1{#1}\fi
\providecommand{\url}[1]{\texttt{#1}}
\providecommand{\href}[2]{#2}
\providecommand{\path}[1]{#1}
\providecommand{\DOIprefix}{doi:}
\providecommand{\ArXivprefix}{arXiv:}
\providecommand{\URLprefix}{URL: }
\providecommand{\Pubmedprefix}{pmid:}
\providecommand{\doi}[1]{\href{http://dx.doi.org/#1}{\path{#1}}}
\providecommand{\Pubmed}[1]{\href{pmid:#1}{\path{#1}}}
\providecommand{\bibinfo}[2]{#2}
\ifx\xfnm\relax \def\xfnm[#1]{\unskip,\space#1}\fi
%Type = Article
\bibitem[{Alder and Wainwright(1959)}]{alder1959studies}
\bibinfo{author}{Alder, B.J.}, \bibinfo{author}{Wainwright, T.E.},
  \bibinfo{year}{1959}.
\newblock \bibinfo{title}{Studies in molecular dynamics. i. general method}.
\newblock \bibinfo{journal}{The Journal of Chemical Physics}
  \bibinfo{volume}{31}, \bibinfo{pages}{459--466}.
%Type = Article
\bibitem[{Ariza et~al.(2012)Ariza, Romero, Ponga and Ortiz}]{ariza2012hotqc}
\bibinfo{author}{Ariza, M.}, \bibinfo{author}{Romero, I.},
  \bibinfo{author}{Ponga, M.}, \bibinfo{author}{Ortiz, M.},
  \bibinfo{year}{2012}.
\newblock \bibinfo{title}{Hotqc simulation of nanovoid growth under tension in
  copper}.
\newblock \bibinfo{journal}{International journal of fracture}
  \bibinfo{volume}{174}, \bibinfo{pages}{75--85}.
%Type = Article
\bibitem[{Bain and Dunkirk(1924)}]{bain1924nature}
\bibinfo{author}{Bain, E.C.}, \bibinfo{author}{Dunkirk, N.},
  \bibinfo{year}{1924}.
\newblock \bibinfo{title}{The nature of martensite}.
\newblock \bibinfo{journal}{trans. AIME} \bibinfo{volume}{70},
  \bibinfo{pages}{25--47}.
%Type = Article
\bibitem[{Batzner et~al.(2022)Batzner, Musaelian, Sun, Geiger, Mailoa,
  Kornbluth, Molinari, Smidt and Kozinsky}]{batzner20223}
\bibinfo{author}{Batzner, S.}, \bibinfo{author}{Musaelian, A.},
  \bibinfo{author}{Sun, L.}, \bibinfo{author}{Geiger, M.},
  \bibinfo{author}{Mailoa, J.P.}, \bibinfo{author}{Kornbluth, M.},
  \bibinfo{author}{Molinari, N.}, \bibinfo{author}{Smidt, T.E.},
  \bibinfo{author}{Kozinsky, B.}, \bibinfo{year}{2022}.
\newblock \bibinfo{title}{E (3)-equivariant graph neural networks for
  data-efficient and accurate interatomic potentials}.
\newblock \bibinfo{journal}{Nature communications} \bibinfo{volume}{13},
  \bibinfo{pages}{1--11}.
%Type = Book
\bibitem[{BELLMAN(1961)}]{COD}
\bibinfo{author}{BELLMAN, R.}, \bibinfo{year}{1961}.
\newblock \bibinfo{title}{Adaptive Control Processes: A Guided Tour}.
\newblock \bibinfo{publisher}{Princeton University Press}.
\newblock \URLprefix \url{http://www.jstor.org/stable/j.ctt183ph6v}.
%Type = Techreport
\bibitem[{Bertsimas et~al.(2006)Bertsimas, Doan and
  Lasserre}]{bertsimas2006multivariate}
\bibinfo{author}{Bertsimas, D.}, \bibinfo{author}{Doan, X.V.},
  \bibinfo{author}{Lasserre, J.}, \bibinfo{year}{2006}.
\newblock \bibinfo{title}{Multivariate exponential integral approximations: a
  moment approach}.
\newblock \bibinfo{type}{Technical Report}. Technical Report, Operations
  Research Center, MIT.
%Type = Article
\bibitem[{Beylkin and Mohlenkamp(2005)}]{beylkin2005algorithms}
\bibinfo{author}{Beylkin, G.}, \bibinfo{author}{Mohlenkamp, M.J.},
  \bibinfo{year}{2005}.
\newblock \bibinfo{title}{Algorithms for numerical analysis in high
  dimensions}.
\newblock \bibinfo{journal}{SIAM Journal on Scientific Computing}
  \bibinfo{volume}{26}, \bibinfo{pages}{2133--2159}.
%Type = Book
\bibitem[{Caflisch et~al.(1997)Caflisch, Morokoff and
  Owen}]{caflisch1997valuation}
\bibinfo{author}{Caflisch, R.E.}, \bibinfo{author}{Morokoff, W.J.},
  \bibinfo{author}{Owen, A.B.}, \bibinfo{year}{1997}.
\newblock \bibinfo{title}{Valuation of mortgage backed securities using
  Brownian bridges to reduce effective dimension}. volume~\bibinfo{volume}{24}.
\newblock \bibinfo{publisher}{Department of Mathematics, University of
  California, Los Angeles}.
%Type = Article
\bibitem[{Chen and Chen(1993)}]{Chen1993}
\bibinfo{author}{Chen, T.}, \bibinfo{author}{Chen, H.}, \bibinfo{year}{1993}.
\newblock \bibinfo{title}{{Approximations of continuous functionals by neural
  networks with application to dynamic systems}}.
\newblock \bibinfo{journal}{IEEE Transactions on Neural Networks}
  \bibinfo{volume}{4}, \bibinfo{pages}{910--918}.
\newblock \URLprefix \url{http://ieeexplore.ieee.org/document/286886/},
  \DOIprefix\doi{10.1109/72.286886}.
%Type = Inproceedings
\bibitem[{Cohen and Welling(2016)}]{Cohen2016}
\bibinfo{author}{Cohen, T.}, \bibinfo{author}{Welling, M.},
  \bibinfo{year}{2016}.
\newblock \bibinfo{title}{Group equivariant convolutional networks}, in:
  \bibinfo{booktitle}{International conference on machine learning},
  \bibinfo{organization}{PMLR}. pp. \bibinfo{pages}{2990--2999}.
%Type = Article
\bibitem[{Dai et~al.(2006)Dai, Kong, Li and Liu}]{dai2006extended}
\bibinfo{author}{Dai, X.}, \bibinfo{author}{Kong, Y.}, \bibinfo{author}{Li,
  J.}, \bibinfo{author}{Liu, B.}, \bibinfo{year}{2006}.
\newblock \bibinfo{title}{Extended finnis--sinclair potential for bcc and fcc
  metals and alloys}.
\newblock \bibinfo{journal}{Journal of Physics: Condensed Matter}
  \bibinfo{volume}{18}, \bibinfo{pages}{4527}.
%Type = Article
\bibitem[{Dongare(2014)}]{dongare2014quasi}
\bibinfo{author}{Dongare, A.M.}, \bibinfo{year}{2014}.
\newblock \bibinfo{title}{Quasi-coarse-grained dynamics: modelling of metallic
  materials at mesoscales}.
\newblock \bibinfo{journal}{Philosophical Magazine} \bibinfo{volume}{94},
  \bibinfo{pages}{3877--3897}.
%Type = Article
\bibitem[{Donoho et~al.(2000)}]{donoho2000high}
\bibinfo{author}{Donoho, D.L.}, et~al., \bibinfo{year}{2000}.
\newblock \bibinfo{title}{High-dimensional data analysis: The curses and
  blessings of dimensionality}.
\newblock \bibinfo{journal}{AMS math challenges lecture} \bibinfo{volume}{1},
  \bibinfo{pages}{32}.
%Type = Inproceedings
\bibitem[{Engel et~al.(2005)Engel, Mannor and Meir}]{engel2005reinforcement}
\bibinfo{author}{Engel, Y.}, \bibinfo{author}{Mannor, S.},
  \bibinfo{author}{Meir, R.}, \bibinfo{year}{2005}.
\newblock \bibinfo{title}{Reinforcement learning with gaussian processes}, in:
  \bibinfo{booktitle}{Proceedings of the 22nd international conference on
  Machine learning}, pp. \bibinfo{pages}{201--208}.
%Type = Book
\bibitem[{Evans and Morriss(2007)}]{j2007statistical}
\bibinfo{author}{Evans, D.J.}, \bibinfo{author}{Morriss, G.P.},
  \bibinfo{year}{2007}.
\newblock \bibinfo{title}{Statistical mechanics of nonequilbrium liquids}.
\newblock \bibinfo{publisher}{ANU Press}.
%Type = Article
\bibitem[{Freitas et~al.(2016)Freitas, Asta and
  De~Koning}]{freitas2016nonequilibrium}
\bibinfo{author}{Freitas, R.}, \bibinfo{author}{Asta, M.},
  \bibinfo{author}{De~Koning, M.}, \bibinfo{year}{2016}.
\newblock \bibinfo{title}{Nonequilibrium free-energy calculation of solids
  using lammps}.
\newblock \bibinfo{journal}{Computational Materials Science}
  \bibinfo{volume}{112}, \bibinfo{pages}{333--341}.
%Type = Book
\bibitem[{Frenkel and Smit(2001)}]{frenkel2001understanding}
\bibinfo{author}{Frenkel, D.}, \bibinfo{author}{Smit, B.},
  \bibinfo{year}{2001}.
\newblock \bibinfo{title}{Understanding molecular simulation: from algorithms
  to applications}. volume~\bibinfo{volume}{1}.
\newblock \bibinfo{publisher}{Elsevier}.
%Type = Incollection
\bibitem[{Garcke et~al.(2010)Garcke, Griebel and Thess}]{garcke2010data}
\bibinfo{author}{Garcke, J.}, \bibinfo{author}{Griebel, M.},
  \bibinfo{author}{Thess, M.}, \bibinfo{year}{2010}.
\newblock \bibinfo{title}{Data mining for the category management in the retail
  market}, in: \bibinfo{booktitle}{Production Factor Mathematics}.
  \bibinfo{publisher}{Springer}, pp. \bibinfo{pages}{81--92}.
%Type = Article
\bibitem[{Gibson et~al.(1960)Gibson, Goland, Milgram and
  Vineyard}]{gibson1960dynamics}
\bibinfo{author}{Gibson, J.}, \bibinfo{author}{Goland, A.N.},
  \bibinfo{author}{Milgram, M.}, \bibinfo{author}{Vineyard, G.},
  \bibinfo{year}{1960}.
\newblock \bibinfo{title}{Dynamics of radiation damage}.
\newblock \bibinfo{journal}{Physical Review} \bibinfo{volume}{120},
  \bibinfo{pages}{1229}.
%Type = Article
\bibitem[{Gorban and Tyukin(2018)}]{gorban2018blessing}
\bibinfo{author}{Gorban, A.N.}, \bibinfo{author}{Tyukin, I.Y.},
  \bibinfo{year}{2018}.
\newblock \bibinfo{title}{Blessing of dimensionality: mathematical foundations
  of the statistical physics of data}.
\newblock \bibinfo{journal}{Philosophical Transactions of the Royal Society A:
  Mathematical, Physical and Engineering Sciences} \bibinfo{volume}{376},
  \bibinfo{pages}{20170237}.
%Type = Book
\bibitem[{Griebel(2005)}]{griebel2005sparse}
\bibinfo{author}{Griebel, M.}, \bibinfo{year}{2005}.
\newblock \bibinfo{title}{Sparse grids and related approximation schemes for
  higher dimensional problems}.
\newblock \bibinfo{publisher}{Citeseer}.
%Type = Article
\bibitem[{Gupta et~al.(2021)Gupta, Ortiz and
  Kochmann}]{gupta2021nonequilibrium}
\bibinfo{author}{Gupta, P.}, \bibinfo{author}{Ortiz, M.},
  \bibinfo{author}{Kochmann, D.M.}, \bibinfo{year}{2021}.
\newblock \bibinfo{title}{Nonequilibrium thermomechanics of gaussian phase
  packet crystals: Application to the quasistatic quasicontinuum method}.
\newblock \bibinfo{journal}{Journal of the Mechanics and Physics of Solids}
  \bibinfo{volume}{153}, \bibinfo{pages}{104495}.
%Type = Article
\bibitem[{Hastings(1970)}]{hastings1970monte}
\bibinfo{author}{Hastings, W.K.}, \bibinfo{year}{1970}.
\newblock \bibinfo{title}{{Monte Carlo sampling methods using Markov chains and
  their applications}}.
\newblock \bibinfo{journal}{Biometrika} \bibinfo{volume}{57},
  \bibinfo{pages}{97--109}.
\newblock \URLprefix \url{https://doi.org/10.1093/biomet/57.1.97},
  \DOIprefix\doi{10.1093/biomet/57.1.97},
  \href{http://arxiv.org/abs/https://academic.oup.com/biomet/article-pdf/57/1/97/23940249/57-1-97.pdf}{\tt
  arXiv:https://academic.oup.com/biomet/article-pdf/57/1/97/23940249/57-1-97.pdf}.
%Type = Article
\bibitem[{Henkelman and J{\'o}nsson(1999)}]{henkelman1999dimer}
\bibinfo{author}{Henkelman, G.}, \bibinfo{author}{J{\'o}nsson, H.},
  \bibinfo{year}{1999}.
\newblock \bibinfo{title}{A dimer method for finding saddle points on high
  dimensional potential surfaces using only first derivatives}.
\newblock \bibinfo{journal}{The Journal of chemical physics}
  \bibinfo{volume}{111}, \bibinfo{pages}{7010--7022}.
%Type = Article
\bibitem[{Henkelman and J{\'o}nsson(2000)}]{henkelman2000improved}
\bibinfo{author}{Henkelman, G.}, \bibinfo{author}{J{\'o}nsson, H.},
  \bibinfo{year}{2000}.
\newblock \bibinfo{title}{Improved tangent estimate in the nudged elastic band
  method for finding minimum energy paths and saddle points}.
\newblock \bibinfo{journal}{The Journal of chemical physics}
  \bibinfo{volume}{113}, \bibinfo{pages}{9978--9985}.
%Type = Book
\bibitem[{Holtz(2010)}]{holtz2010sparse}
\bibinfo{author}{Holtz, M.}, \bibinfo{year}{2010}.
\newblock \bibinfo{title}{Sparse grid quadrature in high dimensions with
  applications in finance and insurance}. volume~\bibinfo{volume}{77}.
\newblock \bibinfo{publisher}{Springer Science \& Business Media}.
%Type = Incollection
\bibitem[{Kainen(1997)}]{kainen1997utilizing}
\bibinfo{author}{Kainen, P.C.}, \bibinfo{year}{1997}.
\newblock \bibinfo{title}{Utilizing geometric anomalies of high dimension: When
  complexity makes computation easier}, in: \bibinfo{booktitle}{Computer
  intensive methods in control and signal processing}.
  \bibinfo{publisher}{Springer}, pp. \bibinfo{pages}{283--294}.
%Type = Inproceedings
\bibitem[{Kim and Yang(2020)}]{kim2020hamilton}
\bibinfo{author}{Kim, J.}, \bibinfo{author}{Yang, I.}, \bibinfo{year}{2020}.
\newblock \bibinfo{title}{Hamilton-jacobi-bellman equations for q-learning in
  continuous time}, in: \bibinfo{booktitle}{Learning for Dynamics and Control},
  \bibinfo{organization}{PMLR}. pp. \bibinfo{pages}{739--748}.
%Type = Phdthesis
\bibitem[{Kulkarni(2007)}]{kulkarni2007coarse}
\bibinfo{author}{Kulkarni, Y.}, \bibinfo{year}{2007}.
\newblock \bibinfo{title}{Coarse-graining of atomistic description at finite
  temperature}.
\newblock Ph.D. thesis. California Institute of Technology.
%Type = Article
\bibitem[{Kulkarni et~al.(2008)Kulkarni, Knap and
  Ortiz}]{kulkarni2008variational}
\bibinfo{author}{Kulkarni, Y.}, \bibinfo{author}{Knap, J.},
  \bibinfo{author}{Ortiz, M.}, \bibinfo{year}{2008}.
\newblock \bibinfo{title}{A variational approach to coarse graining of
  equilibrium and non-equilibrium atomistic description at finite temperature}.
\newblock \bibinfo{journal}{Journal of the Mechanics and Physics of Solids}
  \bibinfo{volume}{56}, \bibinfo{pages}{1417--1449}.
%Type = Article
\bibitem[{Lanzara et~al.(2019)Lanzara, Maz’ya and Schmidt}]{lanzara2019fast}
\bibinfo{author}{Lanzara, F.}, \bibinfo{author}{Maz’ya, V.},
  \bibinfo{author}{Schmidt, G.}, \bibinfo{year}{2019}.
\newblock \bibinfo{title}{A fast solution method for time dependent
  multidimensional schr{\"o}dinger equations}.
\newblock \bibinfo{journal}{Applicable Analysis} \bibinfo{volume}{98},
  \bibinfo{pages}{408--429}.
%Type = Book
\bibitem[{Lecca and Re(2019)}]{lecca2019theoretical}
\bibinfo{author}{Lecca, P.}, \bibinfo{author}{Re, A.}, \bibinfo{year}{2019}.
\newblock \bibinfo{title}{Theoretical physics for biological systems}.
\newblock \bibinfo{publisher}{CRC Press}.
%Type = Article
\bibitem[{LeCun et~al.(2015)LeCun, Bengio and Hinton}]{LeCun2015}
\bibinfo{author}{LeCun, Y.}, \bibinfo{author}{Bengio, Y.},
  \bibinfo{author}{Hinton, G.}, \bibinfo{year}{2015}.
\newblock \bibinfo{title}{{Deep learning}}.
\newblock \bibinfo{journal}{Nature} \bibinfo{volume}{521},
  \bibinfo{pages}{436--444}.
\newblock \URLprefix \url{http://www.nature.com/articles/nature14539},
  \DOIprefix\doi{10.1038/nature14539}.
%Type = Article
\bibitem[{Lee et~al.(2012)Lee, Baskes, Valone and Doll}]{lee2012atomistic}
\bibinfo{author}{Lee, T.}, \bibinfo{author}{Baskes, M.I.},
  \bibinfo{author}{Valone, S.M.}, \bibinfo{author}{Doll, J.},
  \bibinfo{year}{2012}.
\newblock \bibinfo{title}{Atomistic modeling of thermodynamic equilibrium and
  polymorphism of iron}.
\newblock \bibinfo{journal}{Journal of Physics: Condensed Matter}
  \bibinfo{volume}{24}, \bibinfo{pages}{225404}.
%Type = Article
\bibitem[{Li et~al.(2011)Li, Sarkar, Cox, Lenosky, Bitzek and
  Wang}]{li2011diffusive}
\bibinfo{author}{Li, J.}, \bibinfo{author}{Sarkar, S.}, \bibinfo{author}{Cox,
  W.T.}, \bibinfo{author}{Lenosky, T.J.}, \bibinfo{author}{Bitzek, E.},
  \bibinfo{author}{Wang, Y.}, \bibinfo{year}{2011}.
\newblock \bibinfo{title}{Diffusive molecular dynamics and its application to
  nanoindentation and sintering}.
\newblock \bibinfo{journal}{Physical Review B} \bibinfo{volume}{84},
  \bibinfo{pages}{054103}.
%Type = Article
\bibitem[{L{\"o}tstedt and Ferm(2006)}]{lotstedt2006dimensional}
\bibinfo{author}{L{\"o}tstedt, P.}, \bibinfo{author}{Ferm, L.},
  \bibinfo{year}{2006}.
\newblock \bibinfo{title}{Dimensional reduction of the fokker--planck equation
  for stochastic chemical reactions}.
\newblock \bibinfo{journal}{Multiscale Modeling \& Simulation}
  \bibinfo{volume}{5}, \bibinfo{pages}{593--614}.
%Type = Article
\bibitem[{Lu et~al.(2021)Lu, Jin, Pang, Zhang and Karniadakis}]{Lu2021}
\bibinfo{author}{Lu, L.}, \bibinfo{author}{Jin, P.}, \bibinfo{author}{Pang,
  G.}, \bibinfo{author}{Zhang, Z.}, \bibinfo{author}{Karniadakis, G.E.},
  \bibinfo{year}{2021}.
\newblock \bibinfo{title}{{Learning nonlinear operators via DeepONet based on
  the universal approximation theorem of operators}}.
\newblock \bibinfo{journal}{Nature Machine Intelligence} \bibinfo{volume}{3},
  \bibinfo{pages}{218--229}.
\newblock \URLprefix \url{https://www.nature.com/articles/s42256-021-00302-5},
  \DOIprefix\doi{10.1038/s42256-021-00302-5}.
%Type = Article
\bibitem[{Ma et~al.(1993)Ma, Hsu and Straub}]{ma1993approximate}
\bibinfo{author}{Ma, J.}, \bibinfo{author}{Hsu, D.}, \bibinfo{author}{Straub,
  J.E.}, \bibinfo{year}{1993}.
\newblock \bibinfo{title}{Approximate solution of the classical liouville
  equation using gaussian phase packet dynamics: Application to enhanced
  equilibrium averaging and global optimization}.
\newblock \bibinfo{journal}{The Journal of chemical physics}
  \bibinfo{volume}{99}, \bibinfo{pages}{4024--4035}.
%Type = Article
\bibitem[{Meiser and Urbassek(2020)}]{meiser2020alpha}
\bibinfo{author}{Meiser, J.}, \bibinfo{author}{Urbassek, H.M.},
  \bibinfo{year}{2020}.
\newblock \bibinfo{title}{$\alpha \leftrightarrow$ $\gamma$ phase
  transformation in iron: comparative study of the influence of the interatomic
  interaction potential}.
\newblock \bibinfo{journal}{Modelling and Simulation in Materials Science and
  Engineering} \bibinfo{volume}{28}, \bibinfo{pages}{055011}.
%Type = Article
\bibitem[{Mendez et~al.(2018)Mendez, Ponga and Ortiz}]{mendez2018diffusive}
\bibinfo{author}{Mendez, J.}, \bibinfo{author}{Ponga, M.},
  \bibinfo{author}{Ortiz, M.}, \bibinfo{year}{2018}.
\newblock \bibinfo{title}{Diffusive molecular dynamics simulations of
  lithiation of silicon nanopillars}.
\newblock \bibinfo{journal}{Journal of the Mechanics and Physics of Solids}
  \bibinfo{volume}{115}, \bibinfo{pages}{123--141}.
%Type = Article
\bibitem[{Meyer and Entel(1998)}]{meyer1998martensite}
\bibinfo{author}{Meyer, R.}, \bibinfo{author}{Entel, P.}, \bibinfo{year}{1998}.
\newblock \bibinfo{title}{Martensite-austenite transition and phonon dispersion
  curves of fe 1- x ni x studied by molecular-dynamics simulations}.
\newblock \bibinfo{journal}{Physical Review B} \bibinfo{volume}{57},
  \bibinfo{pages}{5140}.
%Type = Article
\bibitem[{Mhaskar(2004)}]{mhaskar2004tractability}
\bibinfo{author}{Mhaskar, H.N.}, \bibinfo{year}{2004}.
\newblock \bibinfo{title}{On the tractability of multivariate integration and
  approximation by neural networks}.
\newblock \bibinfo{journal}{Journal of Complexity} \bibinfo{volume}{20},
  \bibinfo{pages}{561--590}.
%Type = Article
\bibitem[{Miao et~al.(2015)Miao, Feher and McCammon}]{Gromacs}
\bibinfo{author}{Miao, Y.}, \bibinfo{author}{Feher, V.},
  \bibinfo{author}{McCammon, J.}, \bibinfo{year}{2015}.
\newblock \bibinfo{title}{Gaussian accelerated molecular dynamics:
  Unconstrained enhanced sampling and free energy calculation}.
\newblock \bibinfo{journal}{Journal of chemical theory and computation}
  \bibinfo{volume}{11}, \bibinfo{pages}{3584--3595}.
\newblock \DOIprefix\doi{10.1021/acs.jctc.5b00436}.
%Type = Article
\bibitem[{Mishin(2021)}]{mishin2021machine}
\bibinfo{author}{Mishin, Y.}, \bibinfo{year}{2021}.
\newblock \bibinfo{title}{Machine-learning interatomic potentials for materials
  science}.
\newblock \bibinfo{journal}{Acta Materialia} \bibinfo{volume}{214},
  \bibinfo{pages}{116980}.
%Type = Article
\bibitem[{Mishin et~al.(1999)Mishin, Farkas, Mehl and
  Papaconstantopoulos}]{mishin1999interatomic}
\bibinfo{author}{Mishin, Y.}, \bibinfo{author}{Farkas, D.},
  \bibinfo{author}{Mehl, M.}, \bibinfo{author}{Papaconstantopoulos, D.},
  \bibinfo{year}{1999}.
\newblock \bibinfo{title}{Interatomic potentials for monoatomic metals from
  experimental data and ab initio calculations}.
\newblock \bibinfo{journal}{Physical Review B} \bibinfo{volume}{59},
  \bibinfo{pages}{3393}.
%Type = Article
\bibitem[{Morokoff and Caflisch(1995)}]{morokoff1995quasi}
\bibinfo{author}{Morokoff, W.J.}, \bibinfo{author}{Caflisch, R.E.},
  \bibinfo{year}{1995}.
\newblock \bibinfo{title}{Quasi-monte carlo integration}.
\newblock \bibinfo{journal}{Journal of computational physics}
  \bibinfo{volume}{122}, \bibinfo{pages}{218--230}.
%Type = Article
\bibitem[{M{\"u}ller et~al.(2007)M{\"u}ller, Erhart and
  Albe}]{muller2007analytic}
\bibinfo{author}{M{\"u}ller, M.}, \bibinfo{author}{Erhart, P.},
  \bibinfo{author}{Albe, K.}, \bibinfo{year}{2007}.
\newblock \bibinfo{title}{Analytic bond-order potential for bcc and fcc
  iron—comparison with established embedded-atom method potentials}.
\newblock \bibinfo{journal}{Journal of Physics: Condensed Matter}
  \bibinfo{volume}{19}, \bibinfo{pages}{326220}.
%Type = Book
\bibitem[{Niederreiter(1992)}]{niederreiter1992random}
\bibinfo{author}{Niederreiter, H.}, \bibinfo{year}{1992}.
\newblock \bibinfo{title}{Random number generation and quasi-Monte Carlo
  methods}.
\newblock \bibinfo{publisher}{SIAM}.
%Type = Article
\bibitem[{Nos{\'e}(1984)}]{nose1984unified}
\bibinfo{author}{Nos{\'e}, S.}, \bibinfo{year}{1984}.
\newblock \bibinfo{title}{A unified formulation of the constant temperature
  molecular dynamics methods}.
\newblock \bibinfo{journal}{The Journal of chemical physics}
  \bibinfo{volume}{81}, \bibinfo{pages}{511--519}.
%Type = Article
\bibitem[{Onat et~al.(2020)Onat, Ortner and Kermode}]{onat2020sensitivity}
\bibinfo{author}{Onat, B.}, \bibinfo{author}{Ortner, C.},
  \bibinfo{author}{Kermode, J.R.}, \bibinfo{year}{2020}.
\newblock \bibinfo{title}{Sensitivity and dimensionality of atomic environment
  representations used for machine learning interatomic potentials}.
\newblock \bibinfo{journal}{The Journal of Chemical Physics}
  \bibinfo{volume}{153}, \bibinfo{pages}{144106}.
%Type = Article
\bibitem[{Parrinello and Rahman(1981)}]{parrinello1981polymorphic}
\bibinfo{author}{Parrinello, M.}, \bibinfo{author}{Rahman, A.},
  \bibinfo{year}{1981}.
\newblock \bibinfo{title}{Polymorphic transitions in single crystals: A new
  molecular dynamics method}.
\newblock \bibinfo{journal}{Journal of Applied physics} \bibinfo{volume}{52},
  \bibinfo{pages}{7182--7190}.
%Type = Article
\bibitem[{Paskov and Traub(1995)}]{paskov1996faster}
\bibinfo{author}{Paskov, S.}, \bibinfo{author}{Traub, J.},
  \bibinfo{year}{1995}.
\newblock \bibinfo{title}{Faster valuation of financial derivatives}.
\newblock \bibinfo{journal}{J. Portfol. Manage.} .
%Type = Article
\bibitem[{Pestov(2013)}]{pestov2013k}
\bibinfo{author}{Pestov, V.}, \bibinfo{year}{2013}.
\newblock \bibinfo{title}{Is the k-nn classifier in high dimensions affected by
  the curse of dimensionality?}
\newblock \bibinfo{journal}{Computers \& Mathematics with Applications}
  \bibinfo{volume}{65}, \bibinfo{pages}{1427--1437}.
%Type = Article
\bibitem[{Rahman(1964)}]{rahman1964correlations}
\bibinfo{author}{Rahman, A.}, \bibinfo{year}{1964}.
\newblock \bibinfo{title}{Correlations in the motion of atoms in liquid argon}.
\newblock \bibinfo{journal}{Physical review} \bibinfo{volume}{136},
  \bibinfo{pages}{A405}.
%Type = Article
\bibitem[{Sandoval et~al.(2009)Sandoval, Urbassek and Entel}]{sandoval2009bain}
\bibinfo{author}{Sandoval, L.}, \bibinfo{author}{Urbassek, H.M.},
  \bibinfo{author}{Entel, P.}, \bibinfo{year}{2009}.
\newblock \bibinfo{title}{The bain versus nishiyama--wassermann path in the
  martensitic transformation of fe}.
\newblock \bibinfo{journal}{New Journal of Physics} \bibinfo{volume}{11},
  \bibinfo{pages}{103027}.
%Type = Article
\bibitem[{Saxena et~al.(2022)Saxena, Spinola, Gupta and
  Kochmann}]{saxena2022fast}
\bibinfo{author}{Saxena, S.}, \bibinfo{author}{Spinola, M.},
  \bibinfo{author}{Gupta, P.}, \bibinfo{author}{Kochmann, D.M.},
  \bibinfo{year}{2022}.
\newblock \bibinfo{title}{A fast atomistic approach to finite-temperature
  surface elasticity of crystalline solids}.
\newblock \bibinfo{journal}{Computational Materials Science}
  \bibinfo{volume}{211}, \bibinfo{pages}{111511}.
\newblock \DOIprefix\doi{https://doi.org/10.1016/j.commatsci.2022.111511}.
%Type = Article
\bibitem[{Sch{\"u}tt et~al.(2017a)Sch{\"u}tt, Kindermans, Sauceda~Felix,
  Chmiela, Tkatchenko and M{\"u}ller}]{Schutt2017}
\bibinfo{author}{Sch{\"u}tt, K.}, \bibinfo{author}{Kindermans, P.J.},
  \bibinfo{author}{Sauceda~Felix, H.E.}, \bibinfo{author}{Chmiela, S.},
  \bibinfo{author}{Tkatchenko, A.}, \bibinfo{author}{M{\"u}ller, K.R.},
  \bibinfo{year}{2017}a.
\newblock \bibinfo{title}{Schnet: A continuous-filter convolutional neural
  network for modeling quantum interactions}.
\newblock \bibinfo{journal}{Advances in neural information processing systems}
  \bibinfo{volume}{30}.
%Type = Article
\bibitem[{Sch{\"u}tt et~al.(2017b)Sch{\"u}tt, Arbabzadah, Chmiela, M{\"u}ller
  and Tkatchenko}]{schutt2017quantum}
\bibinfo{author}{Sch{\"u}tt, K.T.}, \bibinfo{author}{Arbabzadah, F.},
  \bibinfo{author}{Chmiela, S.}, \bibinfo{author}{M{\"u}ller, K.R.},
  \bibinfo{author}{Tkatchenko, A.}, \bibinfo{year}{2017}b.
\newblock \bibinfo{title}{Quantum-chemical insights from deep tensor neural
  networks}.
\newblock \bibinfo{journal}{Nature communications} \bibinfo{volume}{8},
  \bibinfo{pages}{1--8}.
%Type = Phdthesis
\bibitem[{Sj{\"o}berg(2005)}]{sjoberg2005numerical}
\bibinfo{author}{Sj{\"o}berg, P.}, \bibinfo{year}{2005}.
\newblock \bibinfo{title}{Numerical solution of the Fokker--Planck
  approximation of the chemical master equation}.
\newblock Ph.D. thesis. Uppsala University.
%Type = Article
\bibitem[{Sloan et~al.(2004)Sloan, Wang and Wo{\'z}niakowski}]{sloan2004finite}
\bibinfo{author}{Sloan, I.H.}, \bibinfo{author}{Wang, X.},
  \bibinfo{author}{Wo{\'z}niakowski, H.}, \bibinfo{year}{2004}.
\newblock \bibinfo{title}{Finite-order weights imply tractability of
  multivariate integration}.
\newblock \bibinfo{journal}{Journal of Complexity} \bibinfo{volume}{20},
  \bibinfo{pages}{46--74}.
%Type = Incollection
\bibitem[{Sloan and Wo{\'z}niakowski(2004)}]{sloan2004does}
\bibinfo{author}{Sloan, I.H.}, \bibinfo{author}{Wo{\'z}niakowski, H.},
  \bibinfo{year}{2004}.
\newblock \bibinfo{title}{When does monte carlo depend polynomially on the
  number of variables?}, in: \bibinfo{booktitle}{Monte Carlo and Quasi-Monte
  Carlo Methods 2002}. \bibinfo{publisher}{Springer}, pp.
  \bibinfo{pages}{407--437}.
%Type = Book
\bibitem[{Stroud(1971)}]{stroud1971approximate}
\bibinfo{author}{Stroud, A.H.}, \bibinfo{year}{1971}.
\newblock \bibinfo{title}{{Approximate calculation of multiple integrals}}.
\newblock Prentice-Hall series in automatic computation,
  \bibinfo{publisher}{Prentice-Hall}, \bibinfo{address}{Englewood Cliffs, NJ}.
\newblock \URLprefix \url{https://cds.cern.ch/record/104291}.
%Type = Book
\bibitem[{Sullivan(2015)}]{sullivan2015introduction}
\bibinfo{author}{Sullivan, T.J.}, \bibinfo{year}{2015}.
\newblock \bibinfo{title}{Introduction to uncertainty quantification}.
  volume~\bibinfo{volume}{63}.
\newblock \bibinfo{publisher}{Springer}.
%Type = Article
\bibitem[{Tadmor et~al.(2013)Tadmor, Legoll, Kim, Dupuy and
  Miller}]{tadmor2013finite}
\bibinfo{author}{Tadmor, E.B.}, \bibinfo{author}{Legoll, F.},
  \bibinfo{author}{Kim, W.}, \bibinfo{author}{Dupuy, L.},
  \bibinfo{author}{Miller, R.E.}, \bibinfo{year}{2013}.
\newblock \bibinfo{title}{Finite-temperature quasi-continuum}.
\newblock \bibinfo{journal}{Applied Mechanics Reviews} \bibinfo{volume}{65}.
%Type = Book
\bibitem[{Tadmor and Miller(2011)}]{tadmor2011modeling}
\bibinfo{author}{Tadmor, E.B.}, \bibinfo{author}{Miller, R.E.},
  \bibinfo{year}{2011}.
\newblock \bibinfo{title}{Modeling materials: continuum, atomistic and
  multiscale techniques}.
\newblock \bibinfo{publisher}{Cambridge University Press}.
%Type = Book
\bibitem[{Taylor(2019)}]{taylor2019applications}
\bibinfo{author}{Taylor, C.}, \bibinfo{year}{2019}.
\newblock \bibinfo{title}{Applications Of Dynamic Programming To Agricultural
  Decision Problems}.
\newblock \bibinfo{publisher}{CRC Press}.
\newblock \URLprefix \url{https://books.google.ch/books?id=71SsDwAAQBAJ}.
%Type = Phdthesis
\bibitem[{Tembhekar(2018)}]{tembhekar2018fully}
\bibinfo{author}{Tembhekar, I.}, \bibinfo{year}{2018}.
\newblock \bibinfo{title}{The Fully Nonlocal, Finite-Temperature, Adaptive 3D
  Quasicontinuum Method for Bridging Across Scales}.
\newblock Ph.D. thesis. California Institute of Technology.
%Type = Article
\bibitem[{Thomas et~al.(2018a)Thomas, Smidt, Kearnes, Yang, Li, Kohlhoff and
  Riley}]{thomas2018tensor}
\bibinfo{author}{Thomas, N.}, \bibinfo{author}{Smidt, T.},
  \bibinfo{author}{Kearnes, S.}, \bibinfo{author}{Yang, L.},
  \bibinfo{author}{Li, L.}, \bibinfo{author}{Kohlhoff, K.},
  \bibinfo{author}{Riley, P.}, \bibinfo{year}{2018}a.
\newblock \bibinfo{title}{Tensor field networks: Rotation-and
  translation-equivariant neural networks for 3d point clouds}.
\newblock \bibinfo{journal}{arXiv preprint arXiv:1802.08219} .
%Type = Article
\bibitem[{Thomas et~al.(2018b)Thomas, Smidt, Kearnes, Yang, Li, Kohlhoff and
  Riley}]{Thomas2018}
\bibinfo{author}{Thomas, N.}, \bibinfo{author}{Smidt, T.E.},
  \bibinfo{author}{Kearnes, S.M.}, \bibinfo{author}{Yang, L.},
  \bibinfo{author}{Li, L.}, \bibinfo{author}{Kohlhoff, K.},
  \bibinfo{author}{Riley, P.F.}, \bibinfo{year}{2018}b.
\newblock \bibinfo{title}{Tensor field networks: Rotation- and
  translation-equivariant neural networks for 3d point clouds}.
\newblock \bibinfo{journal}{ArXiv} \bibinfo{volume}{abs/1802.08219}.
%Type = Article
\bibitem[{Thompson et~al.(2022)Thompson, Aktulga, Berger, Bolintineanu, Brown,
  Crozier, in~'t Veld, Kohlmeyer, Moore, Nguyen, Shan, Stevens, Tranchida,
  Trott and Plimpton}]{LAMMPS}
\bibinfo{author}{Thompson, A.P.}, \bibinfo{author}{Aktulga, H.M.},
  \bibinfo{author}{Berger, R.}, \bibinfo{author}{Bolintineanu, D.S.},
  \bibinfo{author}{Brown, W.M.}, \bibinfo{author}{Crozier, P.S.},
  \bibinfo{author}{in~'t Veld, P.J.}, \bibinfo{author}{Kohlmeyer, A.},
  \bibinfo{author}{Moore, S.G.}, \bibinfo{author}{Nguyen, T.D.},
  \bibinfo{author}{Shan, R.}, \bibinfo{author}{Stevens, M.J.},
  \bibinfo{author}{Tranchida, J.}, \bibinfo{author}{Trott, C.},
  \bibinfo{author}{Plimpton, S.J.}, \bibinfo{year}{2022}.
\newblock \bibinfo{title}{{LAMMPS} - a flexible simulation tool for
  particle-based materials modeling at the atomic, meso, and continuum scales}.
\newblock \bibinfo{journal}{Comp. Phys. Comm.} \bibinfo{volume}{271},
  \bibinfo{pages}{108171}.
\newblock \DOIprefix\doi{10.1016/j.cpc.2021.108171}.
%Type = Article
\bibitem[{Tuffin(2004)}]{tuffin2004randomization}
\bibinfo{author}{Tuffin, B.}, \bibinfo{year}{2004}.
\newblock \bibinfo{title}{Randomization of quasi-monte carlo methods for error
  estimation: Survey and normal approximation}.
\newblock \bibinfo{journal}{Monte Carlo Methods and Applications}
  \bibinfo{volume}{10}, \bibinfo{pages}{617--628}.
\newblock \DOIprefix\doi{10.1515/mcma.2004.10.3-4.617}.
%Type = Article
\bibitem[{Venkiteswaran and Junk(2005a)}]{venkiteswaran2005qmc}
\bibinfo{author}{Venkiteswaran, G.}, \bibinfo{author}{Junk, M.},
  \bibinfo{year}{2005}a.
\newblock \bibinfo{title}{A qmc approach for high dimensional fokker--planck
  equations modelling polymeric liquids}.
\newblock \bibinfo{journal}{Mathematics and Computers in Simulation}
  \bibinfo{volume}{68}, \bibinfo{pages}{43--56}.
%Type = Article
\bibitem[{Venkiteswaran and Junk(2005b)}]{venkiteswaran2005quasi}
\bibinfo{author}{Venkiteswaran, G.}, \bibinfo{author}{Junk, M.},
  \bibinfo{year}{2005}b.
\newblock \bibinfo{title}{Quasi-monte carlo algorithms for diffusion equations
  in high dimensions}.
\newblock \bibinfo{journal}{Mathematics and Computers in Simulation}
  \bibinfo{volume}{68}, \bibinfo{pages}{23--41}.
%Type = Article
\bibitem[{Voter(1997)}]{voter1997method}
\bibinfo{author}{Voter, A.F.}, \bibinfo{year}{1997}.
\newblock \bibinfo{title}{A method for accelerating the molecular dynamics
  simulation of infrequent events}.
\newblock \bibinfo{journal}{The Journal of chemical physics}
  \bibinfo{volume}{106}, \bibinfo{pages}{4665--4677}.
%Type = Article
\bibitem[{Voter(1998)}]{voter1998parallel}
\bibinfo{author}{Voter, A.F.}, \bibinfo{year}{1998}.
\newblock \bibinfo{title}{Parallel replica method for dynamics of infrequent
  events}.
\newblock \bibinfo{journal}{Physical Review B} \bibinfo{volume}{57},
  \bibinfo{pages}{R13985}.
%Type = Article
\bibitem[{Wang(2003)}]{wang2003strong}
\bibinfo{author}{Wang, X.}, \bibinfo{year}{2003}.
\newblock \bibinfo{title}{Strong tractability of multivariate integration using
  quasi--monte carlo algorithms}.
\newblock \bibinfo{journal}{Mathematics of Computation} \bibinfo{volume}{72},
  \bibinfo{pages}{823--838}.
%Type = Article
\bibitem[{Wang and Sloan(2005)}]{wang2005high}
\bibinfo{author}{Wang, X.}, \bibinfo{author}{Sloan, I.H.},
  \bibinfo{year}{2005}.
\newblock \bibinfo{title}{Why are high-dimensional finance problems often of
  low effective dimension?}
\newblock \bibinfo{journal}{SIAM Journal on Scientific Computing}
  \bibinfo{volume}{27}, \bibinfo{pages}{159--183}.
%Type = Inproceedings
\bibitem[{Worrall et~al.(2017)Worrall, Garbin, Turmukhambetov and
  Brostow}]{worrall2017harmonic}
\bibinfo{author}{Worrall, D.E.}, \bibinfo{author}{Garbin, S.J.},
  \bibinfo{author}{Turmukhambetov, D.}, \bibinfo{author}{Brostow, G.J.},
  \bibinfo{year}{2017}.
\newblock \bibinfo{title}{Harmonic networks: Deep translation and rotation
  equivariance}, in: \bibinfo{booktitle}{Proceedings of the IEEE Conference on
  Computer Vision and Pattern Recognition}, pp. \bibinfo{pages}{5028--5037}.
%Type = Article
\bibitem[{Zhou et~al.(2020)Zhou, Cui, Hu, Zhang, Yang, Liu, Wang, Li and
  Sun}]{Zhou2020}
\bibinfo{author}{Zhou, J.}, \bibinfo{author}{Cui, G.}, \bibinfo{author}{Hu,
  S.}, \bibinfo{author}{Zhang, Z.}, \bibinfo{author}{Yang, C.},
  \bibinfo{author}{Liu, Z.}, \bibinfo{author}{Wang, L.}, \bibinfo{author}{Li,
  C.}, \bibinfo{author}{Sun, M.}, \bibinfo{year}{2020}.
\newblock \bibinfo{title}{{Graph neural networks: A review of methods and
  applications}}.
\newblock \bibinfo{journal}{AI Open} \bibinfo{volume}{1},
  \bibinfo{pages}{57--81}.
\newblock \URLprefix
  \url{https://linkinghub.elsevier.com/retrieve/pii/S2666651021000012},
  \DOIprefix\doi{10.1016/j.aiopen.2021.01.001}.
%Type = Book
\bibitem[{Ziman(2001)}]{ziman2001electrons}
\bibinfo{author}{Ziman, J.M.}, \bibinfo{year}{2001}.
\newblock \bibinfo{title}{Electrons and phonons: the theory of transport
  phenomena in solids}.
\newblock \bibinfo{publisher}{{Oxford University Press}},
  \bibinfo{address}{Oxford, UK}.
%Type = Article
\bibitem[{Zubarev and Morozov(1989)}]{zubarev1989nonequilibrium}
\bibinfo{author}{Zubarev, D.N.}, \bibinfo{author}{Morozov, V.G.},
  \bibinfo{year}{1989}.
\newblock \bibinfo{title}{Nonequilibrium statistical ensembles in kinetic
  theory and hydrodynamics}.
\newblock \bibinfo{journal}{IN: Statistical mechanics and theory of dynamic
  systems-In celebration of the 80th birthday of Academician Nikolai
  Nikolaevich Bogoliubov (A91-29994 11-77), Moscow} ,
  \bibinfo{pages}{140--151}.

\end{thebibliography}

\end{document}